# AHA! Strategies for Gaining Insights into Software Design

MARY SHAW, Carnegie Mellon University, USA, mary.shaw@cs.cmu.edu

Shepherd: Richard P Gabriel, poet, writer, computer scientist, California

These patterns describes the strategies I use to find novel or unorthodox insights in the area of software design and research. The patterns are driven by inconsistencies between what we say and what we do, and they provide techniques for finding actionable insights to address these inconsistencies. These insights may help to identify research opportunities; they may stimulate critiques of either research or practice; they may suggest new methods.

CCS Concepts • **Software and its engineering** • Software creation and management • Designing software

Additional Keywords and Phrases: software design, software engineering, software architecture, software research topics, software innovation

## 1 MOTIVATION

These **Aha!** patterns describe my strategies for finding novel or innovative insights into software design and programming languages. They document the strategies I use to identify and respond to discrepancies between doctrine and practice and to frame research questions that address these discrepancies.

The key idea is noticing dissonance: inconsistency between the way we talk about ideas and actual software practice. Insights arise from a dialog between the designer and the sources of dissonance, possibly extending over years: First, there's reflection on a significant discrepancy between what we commonly say we do (theory) and what we actually do (practice); this discrepancy may be deterring us from addressing a practical problem. This dissonance may lead to a response that challenges a popular assumption about theory or practice, or it may trigger a reframing of the situation, for example by introducing an unorthodox element suggested by some other field. That response may be flawed or incomplete, which may lead in turn to reflection on the response and further refinement. In Donald Schön's terms [Schön 1983], this is a reflective conversation with the situation.

Thus there are two types of patterns: one for the ways dissonance triggers questions, and five for ways to seek novel or unorthodox insights. The first type identifies inconsistencies between our words and actions; the second type identifies ways to see the inconsistencies from new points of view that may avoid or resolve the dissonance.

The pattern for triggering questions is:

> **Dissonance**: Notice inconsistencies between what we say and what we do. (p.2)

Five patterns for seeking insights are:

> **Satisfice**: Accept "good enough" as good enough. (p.7)
> **Reframe:** Reinterpret the situation from a new point of view. (p.10**)**
> **Classify**: Find underlying structure. (p.14)
> **Import**: Adapt ideas from other fields. (p.17)
> **Satirize**: Ridicule absurdity. (p.22)

# The Aha! Patterns

## 2  PATTERN FOR TRIGGERING QUESTIONS

### Pattern 1.  DISSONANCE: Notice inconsistencies between what we say and what we do

> "When the terrain disagrees with the map, trust the terrain"
>
> — Swiss Army Proverb
>
> "Question authority"
>
> — Timothy Leary, or maybe Socrates
>
> "Do you seriously believe that?"
>
> — The snarky gremlin that sits on my shoulder and calls me out when I just parrot the party line

**Context**

Software development has two aspects—the actual practice of writing software and the principled models of how software is developed. The goal of the former is producing software that satisfies a practical need, and the goal of the latter is improving the former by providing broadly applicable principles and theories that support reasoning about various properties of the software, especially efficiency and correctness (whatever that may be). Each comes in many variants. Understandably, the two aspects are often inconsistent.

It is not uncommon for principled models about software development to be at odds – even wildly at odds — with actual practice, yet the models often shape conventional wisdom. Such conventional wisdom often glosses over problems or treats them as one-off special cases instead of systematic shortcomings. For example, both researchers and developers still set correctness as a principal objective despite the pervasive practical use of software that isn't actually "correct". Disparities might arise because a sound theory has limited application, or a burst of enthusiasm for a new idea ignores its limitations, or an overzealous proponent of a method exaggerates its effectiveness to gain support.

**Dissonance** can also arise from flawed analogies. The 1980's "software factory" movement aimed to improve software practice by treating software creation like an assembly line process. The analogy is faulty, though, because software development is more appropriately mapped to product design than to producing many instances (that would be pressing the CDs on which software was then distributed).

We see this in other fields, as well. Horst Rittel and Melvin Webber notably identified the dissonance between consultants' formal models and problems situated in the world to the planning policy community [Rittel and Webber 1973]. They gave the label "wicked problems" to these complex, situated social problems with multiple stakeholders and no clear solution criteria. This idea has been imported, not always accurately, into software engineering. In his work on semantics, Alfred Korzybski said, "A map is not the territory it represents, but, if correct, it has a similar structure to the territory, which accounts for its usefulness" [Korzybski 1933, p.58]. That is, models are not reality, but abstractions of reality; they are flawed but useful. More recently, Carl Shapiro and Hal Varian confronted the claim that software was such a new technology that the rules of economics did not apply, that we need a new economics for software [Shapiro and Varian 1998]. To the contrary, they say, we don't need a new economics for software, we just need to apply established economics thoughtfully.

When people become invested in an idea, it's hard for them to move on to competing ideas. This isn't specific to software; it's part of the human condition. In studying designers, Nigel Cross found that they can become fixated on early solution ideas, hanging onto these ideas as long as possible even in the face of difficulties and shortcomings [Cross 2007]. In science as well, theories are often patched up with constraints and assumptions designed to exclude cases that challenge the theory; Imre Lakatos' Proofs and Refutations presents a fictional dialog in which students attempt to prove a theorem but repeatedly encounter counterexamples, which they must patch the theory to handle [Lakatos 1976]. Schön's advice on reflection also applies: even if a dissonance is small, it may be a hint of an underlying anomaly; even if a dissonance is striking, it may best be handled by looking beyond the obvious discrepancies to adjust relations of other parts of the system.

These dissonances between theory and practice and between conventional wisdom and practice often point to good opportunities for innovation, or at least to new insights.

This pattern applies broadly. A hint that it may apply is the widespread repetition of absolutist positions with religious fervor.

### *Problem*

*Sometimes conventional wisdom, current claims, popular theories, or unspoken assumptions of the field do not match actual practice or facts on the ground. Adopting such sweeping claims uncritically can distract from addressing actual problems and opportunities to which they do apply, even if they do address some problems. By closing off other avenues for progress, they can even be counterproductive. This can manifest as dissonance to the software developer and is a sign that theory and practice need to be realigned.*

For example, imposing a requirements-first process on a project that needs exploratory programming to understand the requirements would interfere with solving the problem at hand.

**DISSONANCE** comes in different forms:

- **Narrow theory:** Principled models or theories are unrealistic or limited and don't address practical needs. For example,
    - "Correctness is an essential goal", even though practical software isn't "correct".
    - "Formal specifications capture all the objectives", even though some objectives are not expressible in the formal system.
    - "Software dependability is a function only of availability, reliability, safety, and security", even though the specific properties that contribute to reliability depend on context.

  This dissonance surfaces limits of theory when theory addresses only a subset of practice; it encourages more nuanced, realistic models.

- **Naïve assumptions:** Conventional wisdom is exaggerated, or tacit assumptions about practice ignore important aspects of actual practice or principles; making these claims uncritically (especially when they are sweeping generalizations) can be counterproductive if it steers people away from useful techniques. For example,
    - "Objects will solve all our problems", (or subroutine libraries, or …), even though other structures are better matched to many problems.
    - "Programming languages must be fully general" (Turing complete), even though special-purpose languages are widespread and sometimes incomplete.
    - "Bibliometrics are good measures of quality", 'nuff said.

    This dissonance surfaces obliviousness or shallow thinking; it makes space for other useful alternatives and allows navigation of the generality/power tradeoff.

- **Ad hoc practices:** Current practice is ad hoc and consequently complex, risky, or inscrutable and will benefit from structured, even formal models. For example,
    - "A software system is merely a collection of modules linked together", even though software systems include non-code components in complex relationships.
    - "Procedure calls suffice to explain module interactions", even though interactions among components are much richer than simple call-and-return.
    - "Of course programmers will learn formal logic" or category theory, or whatever, even though the majority of practicing professional programmers never studied math at that level.

    This dissonance surfaces limits of practice; it calls for more systematic techniques.

These dissonances often don't make themselves known as a result of direct analysis, but rather by free association or analogy, or by reflecting on how some aspect of a system is out of tune with the rest of the system.

Some types of **Dissonance** come up repeatedly. For example, the tension between whether the primary objective of a design is to be correct or to be "good enough" for practical purposes may arise from a narrow view of correctness as purely functional correctness. These dissonances lead naturally to associations with corresponding insight patterns. In this case, **Satisfice**.

*Solution*

Identify which forms of dissonance are involved. This will help you decide which aspect of the dissonance to address: limitations of theory, naïveté of conventional wisdom, or haphazardness of practice.

Identify the source of the dissonance – the assumption behind the claim or conventional wisdom or practice – and examples of common things or practices that violate it. This often requires analysis such as

- identifying the scoping assumptions of a formalism
- critical reading of the conventional wisdom in order to identify the unspoken assumptions
- finding structure to organize undisciplined practice

If possible, correct the problematic assumptions, or at least call attention to them. More effectively, identify the way the problematic element falls short and frame a research question around the problem that it masks. For example, call out the inability of a purely functional formalism to capture many quality attributes.

**Dissonance** may arise from differences in context and framing of the dissonant positions. For example, in discussions of generative AI, public dismay over biased results or erroneous information delivered with great confidence arises from applying human intuition about correctness, while at the same time the systems may actually satisfy a narrower definition of correctness that's related to the output being statistically similar to the training set.[1] They results of concern may also arise from inappropriate application of the AI tools. Eugenia Cheng uses category theory to untangle some similar examples [Cheng 2023].

*Related Patterns*

**DISSONANCE** is the pattern for identifying inconsistencies that trigger the search for insight. When this pattern identifies dissonance, the other patterns offer ways to find insights.

Dissonance between perfection and utility often calls for **SATISFICING**. Popular methods often come with hard-wired assumptions about solutions, and these may provide opportunities for **REFRAMING**. Sometimes ad hoc solutions reflect lack of systematic organization of relevant knowledge, which calls for **CLASSIFYING**; sometimes they are hacking at problems that have analogs in other fields, and **IMPORT** can supply ideas based on techniques from other areas.

When the dissonance does not resonate with your audience, it can sometimes be made more obvious by taking some assumption to an extreme or manipulating an unexpected parameter. So, if reason fails, **SATIRIZE** can direct attention to the dissonance. Even if the dissonance can't be resolved (often the case for administrative excess), this may suggest coping strategies.

*Examples*

**DISSONANCE** has served me well as a source of research ideas. I described the strategy informally in "Sparking research ideas from the friction between doctrine and reality" [Shaw 2005 Spark]. This article discusses four examples in which I found research problems by noticing that our descriptions of what we do don't match what we actually do. These ideas about noticing dissonance evolved over time, reaching full flower in "Myths and mythconceptions: what does it mean to be a programming language, anyhow?" [Shaw 2021 Myth], which reflects on the richness of actual practice compared to the origin myth of software development ("Professional programmers create software by writing code in sound programming languages to satisfy a given formal specification and verify that the program is correct; software is made by (just) composing program modules.")

The tension between "good" and "best" arises repeatedly, often a victim of sweeping generalization rooted in the origin myth. Notably, advocates of formal methods assert (or did, strongly, in the 1970s and 1980s) that software must be provably correct to be useful and that any change at all in the software requires a fresh proof. But these same colleagues cheerfully used computers that were anything but correct. Further, that concept of correctness depends on fixed formal specifications and analysis showing the consistency of the specifications with the code. In contrast, the specifications for software products at the time were largely statements about extrafunctional properties and the power of the hardware needed to run the applications. This tension led to a pair of examples: The **SATISFICING** example of *Credentials* deals with incomplete specifications, evolving specifications, and information of varying quality. This sets the stage for the **REFRAMING** examples of a *Calculus of Confidence*, which consider how to integrate information of varying quality. This laid the groundwork for **IMPORTING** ideas from evidence based medicine to *Aggregate Research Results*. In a slightly different vein, a movement to elevate quantitative empirical research above other paradigms in software engineering triggered an early **Classification** of ***Software Engineering Research Types***.

---

[1] The current fashion for calling generative AI's errors "hallucinations" is appalling. It anthropomorphizes the software, and it spins the errors as somehow idiosyncratic quirks of the system even when they're objectively incorrect. Moreover, WebMD's advice about hallucinations is "If you or a loved one has hallucinations, go see a doctor." https://www.webmd.com/schizophrenia/what-are-hallucinations

Over-reaching claims often arise from a good idea passing uncritically into popular use. The originator of the idea very likely had a more nuanced understanding than the popular understanding, but the process of snowballing into the-meme-of-the-month has an unfortunate way of stripping nuance, leading to claims that the idea will solve all our problems. My reaction to just such a fixation on correctness triggered the **Satisficing** example of ***Sufficient Correctness***, in which I consider dependability and how to determine that software is good enough for practical purposes. This led into the **Reframing** example of ***Everyday Dependability***, which challenges the notion that software is in either a working state or a failed state in favor of a model that accepts degrees of degradation and accommodates them through ideas **Imported** from biological ***Homeostasis***.

When practice has just grown up without much thoughtful analysis, or when important aspects of the practice have not received systematic attention, there is often opportunity for improvement from simply focusing on the ad hoc aspects of the practice. The vast majority of people who develop software are not professional software developers, yet software engineering pays little attention to these ***Vernacular Programmers***., so **Reframing** our understanding opens opportunities to provide them better tools. Software developers tend to seek solutions depth-first rather than breadth-first, and elevating the visibility of **Classifying** design knowledge as ***Design Spaces*** provides an approach to a better balance. Specifically in the area of adaptive systems, common practice did not address how effectively the systems actually achieved adaptation, and **Importing** ideas from control theory helped to clarify the ***Design Obligations for Control***.

Ill-conceived policies and procedures sometimes call out for **Satire**.

These examples are elaborated in the discussion of the patterns that generated them. The insight patterns and associated examples are

| Pattern | Narrow theory | Naïve assumptions | Ad hoc practice |
|---|---|---|---|
| **Satisfice** | ***Credentials*** (p.8) | ***Sufficient Correctness*** (p.9) | |
| **Reframe** | ***Calculus of Confidence*** (p.11) | ***Everyday Dependability*** (p.12) | ***Vernacular Programmers*** (p.13) |
| **Classify** | ***SE Research Types*** (p.15) | | ***Design Spaces*** (p.16) |
| **Import** | ***Aggregating Research Results*** (p.19) | ***Homeostasis*** (p,19) | ***Design Obligations for Control*** (p.20) |
| **Satirize** | ***Push the envelope*** (p.23) | ***Thinking not Counting*** (p.23) | ***Whimsy*** (p.24) |

In addition, a final, extended, example describes a sequence of pattern applications that addressed ad hoc practice in software system organization. Over a period spanning decades, the sequence of insights led to the discipline of ***Software Architecture***. The final section describes the flow among these ideas and the **Aha!** patterns that triggered many of them. The examples in the ***Software Architecture*** sequence are

| Examples | Pattern |
|---|---|
| ***Alternatives to Objects*** (p.26) | **Reframe** |
| ***Computer Architecture Concepts*** (p.27) | **Import** |
| ***Software Architecture Styles*** (p.28) | **Reframe** |
| ***Boxology Design Space*** (p.29) | **Classify** |
| ***Languages for Software Architecture*** (p. 30) | **Reframe** |
| ***Formality of Software Architecture*** (p.32) | **Satisfice** |

## 3  PATTERNS FOR NEW INSIGHTS

### *Pattern 2.  SATISFICE: Accept "good enough" as good enough*

> "Il meglio è l'inimico del bene"
> ("The perfect is the enemy of the good")
>
> — Voltaire 1770
>
> "Excellence does not require perfection"
>
> — Henry James
>
> "The real economic actor is in fact a satisficer, a person who accepts 'good enough' alternatives, not because less is preferred to more but because there is no choice"
>
> — Herbert Simon
>
> "Everything should be as formal as necessary, but no formaller"
>
> — paraphrasing Albert Einstein

#### *Context*

Idealism is often in tension with practicality. Indeed, the heart of engineering is making cost-effective tradeoffs between alternatives, for example choices for tolerances, safety factors, and reliability. Yet software and programming language research has often emphasized generality at the expense of domain-specific applicability. Formal theory of programming languages is particularly apt to value generality and soundness over utility.

Focusing on complete or optimal results can impede evaluation of partial-coverage techniques. The completeness mindset cries out for measuring the shortfall from completeness, but it may be more appropriate to measure improvement.

Herbert Simon introduced the concept of "satisficing" (a portmanteau of "satisfying" and "sufficing") to describe the way people make decisions in real life [Simon 1956]. In an imperfect world with imperfect information, it's often not possible to fully optimize. So real-life decisions often accept a result that's above some threshold; this is often good enough when practical costs and constraints are considered. In his Nobel Prize acceptance speech Simon observed that "decision makers can satisfice either by finding optimum solutions for a simplified world, or by finding satisfactory solutions for a more realistic world. Neither approach, in general, dominates the other…" [Simon 1979].

Around 1980, when I was deeply involved in program verification, I was dubious about the value of two empirical models, the keystroke model for predicting user time in interactive systems [Card et al 1980] and the COCOMO model for software project planning [Boehm 1981]. My reaction to both was that they were entirely too simple to be correct. What I missed was that they were accurate enough to be useful while being simple enough to use. This little epiphany set me up for some of these examples.

#### *Problem*

*The desire for rigor, completeness, correctness, or optimality can lead to absolutist or impractical demands.*

Formal and precise analyses are great if they're available, but there are two problems in practice: Formal models usually focus on functionality, but the common understanding of correctness depends on many other properties, such as performance or adherence to social norms. Further, there are costs associated with determining precise values of attributes, whether they be requirements or analyses of existing systems, and if the value of this information does not exceed its cost, it's counterproductive to acquire it In practice, these demands are relaxed, though the decision about how much they can reasonably be relaxed in a particular situation remains ad hoc. The challenge is to understand which constraints can be relaxed, and by how much, especially in light of the evolution of the system.

### *Solution*

Recognize the limitations of the strong demands for purity and perfection, for example that

- "proof of correctness" of code really refers to a formal demonstration that the specification is consistent with the code, but the client often views correctness as including properties that can't be stated in the formal system used for proof of correctness, such as quality attributes.
- specifications are supposed to be complete, formal, static, and homogeneous, but in practice that knowledge is partial, approximate, evolving, and heterogeneous.
- there are costs associated with acquiring the information in specifications, which rise as the number of properties and demand for precision increase, so it is often reasonable to choose a cost-effective level of completeness, precision, and rigor.

Assess the real needs of the current setting. There is often a tradeoff between cost and precision, and approximate solutions are often sufficiently good for the task. For example, engineers love linear models. These models lend themselves to simple mathematical operations, and they're often good enough. The real world may also involve higher-order terms, but the additional precision from carrying those in the analysis often doesn't justify the analytic cost. The trick, of course, is knowing when linear models are good enough. So too is it with software.

Seek ways to support the more ambiguous, less precise assurances that are actually achievable, for example through confidence intervals, explicit handling of incompleteness, expectations about error rates, qualitative analysis, and processes for developing confidence.

Buttress this reduced precision and formality with ways to determine the level of confidence that is required for your specific application. After all, an app for finding the show times for local movies has vastly greater tolerance for failure than life- and safety-critical software for applications like heart pacemakers and automated nuclear power plant shutdown.

### *Related Patterns*

When **Satisficing** has identified the limitations of a theory, **Reframing** the situation may open the door to a good-enough model with different assumptions. **Classifying** may organize the available knowledge in a way that allows the limitations to be addressed. **Importing** a point of view from another area where problems have a similar structure may be helpful.

### *Examples*

Example 1.  ***Credentials***

In 1996 I challenged the conventional doctrine that component specifications are sufficient, complete, static, and homogeneous. This doctrine, which is still in circulation, assumes it is possible to identify all the significant properties of a component and that it is feasible to determine the values of all those properties. In practice, though, there is strong **Dissonance** between this and actual practice: what we can actually know about a component is a subset of what it actually does, and our knowledge about a component changes over time with additional analysis and evidence from use.

Instead of arguing with the formal community about the meaning of "specification", I introduced the concept of "credentials" in "Truth vs knowledge: the difference between what a component does and what we *know* it does" [Shaw 1996 Truth] and argued that partial, incremental, extensible, heterogeneous knowledge is often adequate, in other words that we can **SATISFICE**. Credentials could be represented as a property list—a list of *<attribute-value>* pairs—but the sources of the values are so diverse that it's also important to keep track of the confidence we have in those values. Our information comes not only from high-ceremony sources such as verification and testing but also from low-ceremony sources such as reviews, reputation, and reports from the field. Further, the values may be on different measurement scales. Thus, a credential was (initially) a list of *<attribute, value, credibility>* triples. This addresses the restrictiveness of specifications by allowing richer values of more diverse types, but this leads to the question of how to use the richer values. I addressed that by **REFRAMING** the question as finding a ***Calculus of Confidence.***

Example 2.  ***Sufficient Correctness***

A widespread belief in software engineering holds that functional correctness is mandatory. In the real world, though, we usually don't have complete specifications, even complete functional specifications, for all possible cases. Without a specification there is no verification[2], so what do we do to resolve this **DISSONANCE**? For a large class of problems, the question is whether the system is sufficiently dependable for everyday application, in which occasional misbehavior will be noticed and noncatastrophic. So, building on the idea of ***Credentials*** in 2000 we again **SATISFICED** to consider ***Sufficient Correctness,* REFRAMED** to describe models of ***Everyday Dependability,*** and **IMPORTED** ideas about ***Homeostasis*** from biology as a model of self-healing. We presented these ideas in "Sufficient correctness and homeostasis in open resource coalitions: how much can you trust your software system?" [Shaw 2000 Homeostasis] and "An approach to preserving sufficient correctness in open resource coalitions" [Raz and Shaw 2000 SuffCor].

***Sufficient Correctness*** responds to the dissonance between the "gold standard" of verified correctness and the market realities of 2000 when, as was common practice, Windows 2000 was reported to ship with 63,000 "defects" including more than 21,000 "postponed" bugs [Foley 2000]. Common practice also called for assigning (context-independent) severity levels and only shipping with lower-severity defects.

Certainly there are some applications that really, really need to work. But it's very expensive to achieve this, so for other applications, a more pragmatic expectation is in order. Further, the character of software systems themselves has changed. Traditional closed-shop software development in which a single organization controls all the software in a system and all changes to that software has given way to "open resource coalitions". These are dynamically formed, task-specific assemblages that rely on non-code resources and third-party components that may change without notice to the user. To reason about these coalitions, one must consider interactions among the independent resources and dynamic change, so it is vastly more complex than reasoning about closed-shop systems and correspondingly more limited in its assurances.

 Two factors affect our expectations for dependability: what are the consequences of failure and how likely is it that a person will notice a problem and address it? These expectations allow us to visualize a space in which correctness of fully automated processes with catastrophic consequences should be assured, but ordinary day-to-day applications can be good enough for practical purposes, as suggested in Figure 1. Note that as the capability of software increases, as suggested by the arrows, the need for stronger assurances may also increase. The good-enough realm is the setting for **REFRAMING** the expectations in ***Everyday dependability***.

---

[2] Formal verification shows consistency between the code of a program and its formal specification. It does not address whether the specification accurately captures the designer's intention for the software.

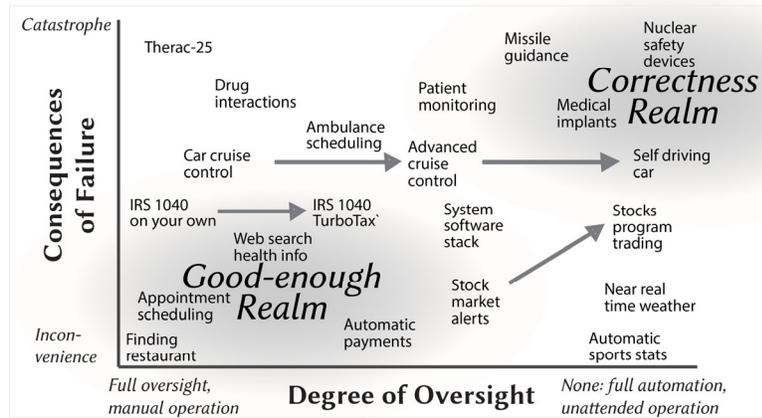

Figure 1. How consequences and human oversight affect dependability requirement.
Originally in [Shaw 2000 Homeostasis] this version from [Shaw 2021 Myth].

### Pattern 3. REFRAME: Reinterpret the situation from a new point of view

"Point of view is worth 80 IQ points"

— Alan Kay, 1982

"All models are wrong, but some are useful"

— George E. P. Box, 1976

#### Context

When program verification was new, we talked about "proving correctness of programs". This was a sweeping exaggeration, because all we could actually do is prove, within certain formal systems, that the code of a program was consistent with its formal specification. Anything in the user's intuition about "correctness" that was not captured in the formal specification was out of bounds. As we began constructing the proofs, we spent about as much time debugging the specifications as we spent debugging the code. Skeptics of verification argued that this showed verification was a waste of time, but some of us recognized the benefit of viewing the software from two points of view; the imperative code and the functional specification. Years later, the multiple notations of UML provided multiple views at a larger scale and higher level of abstraction, though it did not provide a complete set of notations. The REFRAME pattern builds on these observations by encouraging deliberate shifts to different formulations to see the situation in a new light.

Problem framing is an essential step in design. Designers view a frame as an assumption or perspective on a situation. They often address a problem by questioning a client's assumptions, changing them to reframe the situation. Reframing thus becomes a dialog between the technology, the users, and the designers. It is the problem-setting part of Donald Schön's reflective practice: "Problem setting is the process in which, interactively, we *name* the things to which we will attend and *frame* the context in which we will attend to them" [Schön 83]. In *The Reflective Practitioner* he gives an extended example [Schön 1983 pp.191–195] of how experts from six different fields frame a given problem from the viewpoints of their own fields and arrive at very different recommendations for addressing the problem.

Nigel Cross builds on Schön's analysis, noting that experienced architects and expert engineering designers often engage in problem framing—they select problem paradigms or strong guiding themes as generators for setting problem boundaries and goals [Cross 2007].

### *Problem*

*Rigid adherence to a narrow model or uncritical acceptance of conventional wisdom can deter you from exploring a variety of alternatives.*

Defaulting to a familiar solution rather than designing for the problem at hand can lead to misfit solutions. Using a familiar framework will bring the framework's point of view and infrastructure to the problem, whether it's a good match or not. Adhering to a prescribed development method will channel development activities in specific ways. In addition, solving problem depth-first—committing to the first idea that comes along rather than considering a variety of options—can also lead the designer to miss good opportunities.

### *Solution*

Consider the situation from different points of view. Identify tacit assumptions that constrain your view of the situation and consider alternative assumptions, including some contrary ones. Explore several solutions breadth-first, looking at the problem from different points of view.

An important **Reframing** is a shift from perfection to adequacy, a form of **Satisficing**. Asking whether an absolutely optimal schedule is required or it's sufficient to have a schedule that's good enough to stay ahead of demand may vastly simplify the scheduling, though it will require good estimates of demand. Optimality may be vastly more expensive than sufficiency: the full complexity analysis of Euclid's algorithm occupies 17 dense pages of Knuth's *Seminumerical Algorithms* [Knuth 1969 pp.316–333], but the simple observation that one of the two numbers must be halved at each step leads immediately to a bound that's only about a factor of 3 worse [Shaw 1981 Good].

### *Related Patterns*

**Reframing** involves changing the point of view about some aspect of the situation. Accordingly, it can take many forms. **Import** often calls for a type of reframing that looks to another subdiscipline or a different field entirely for inspiration. Other disciplines use different formulations and models, so **Importing** offers a rich source of new points of view. **Classify** is a particular type of reframing that specifically seeks systematic structure, for example design spaces.

### *Examples*

Example 3.  ***Calculus of Confidence***

Reflection on the initial ideas about confidence in ***Credentials*** revealed considerable diversity of evidence, ranging from high-ceremony evidence from formal verification, testing, and careful empirical studies to low-ceremony evidence from reviews, popularity, or qualitative reasoning. Different types of evidence have different properties, and when the evidence is combined, the level of confidence in the result depends the credibility of the individual elements being composed. This creates further **Dissonance** between the ideal for full formal specifications and the pragmatics of real systems.

In addition, even quantitative values for different properties cannot be combined willy-nilly. Values for different properties are heterogeneous—they may arise from different measurement scales (nominal, ordinal, interval, ratio). This limits allowable calculations (for example, 80°F is not "twice as hot" as 40°F, because Fahrenheit is an interval scale, which does not support ratios). Other properties of attributes such as whether they are perishable, fungible, or rival restricts the ways they can be used. In "Time is not money" we showed that ignoring these properties can lead to analysis errors (for example, calendar days and staff months are non-fungible, so schedules cannot be manipulated by changing staffing levels; bandwidth is perishable, so it can't be "saved up" for a future burst of communication) [Poladian et al 2003 TimeMon].

We **Reframed** the *Credentials* idea by considering different kinds of evidence, including uncertain and qualitative evidence, and ways to aggregate it in "Toward a calculus of confidence" [Shaw 2007 Confid]. We recognized that evidence accumulates, and values that may have been derived from specifications of other components must be updated as a result, so in "Developing confidence in software" we added information tracking the source of the value to the representation, making a credential a list of *<attribute, value, credibility, provenance>* quadruples [Scaffidi and Shaw 2007 Cred].

These ideas seeded a later discussion that went beyond combining simple component properties to looking at ways to **Aggregate Research Results** based on an analogy **Imported** from evidence-based medicine [Le Goues 2018 Bridge]. Thus this example began with **Dissonance** triggered by a simplistic assumption about the completeness and formality of specifications, which by **Satisficing** to accepting best available information. This led to a **Reframing** that represented incomplete but "good enough" information, and later to **Importing** ideas from evidence-based medicine about techniques for reasoning with information of varying quality.

Example 4.  ***Everyday dependability***

The mantra of correctness and the naïve view of dependability hold that failures should be prevented. In practice, however, there is **Dissonance** between this view and the practical systems that can recover in various ways from problems.

There are two general ways of dealing with the possibility of bad things happening: Prevent them from happening at all, and detect problems and react to them as they occur. We approach the former through validation and the latter through remediation. In "Strategies for achieving dependability in coalitions of systems" [Shaw 2007 Depend] I organized strategies for validation and remediation as indicated in Figure 2. These distinctions allow dependability problems to be **Reframed** to appropriately match their settings, and they recognize non-technical alternatives such as insurance that remediates via economic compensation.

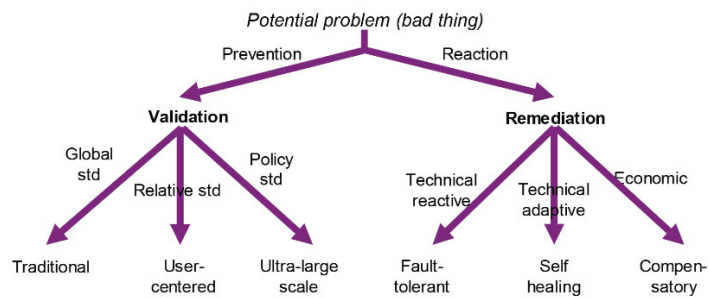

Figure 2.   Techniques for handling dependability problems [Shaw 2007 Depend]

In "Everyday dependability for everyday needs" [Shaw 2002 Everyday] I examined how the shift from closed software systems to open resource coalitions opened new approaches to dependability, in particular to systems that could respond to problems automatically. The remediation branch is appropriate for everyday dependability, as it can tolerate failures. In "An approach to preserving sufficient correctness in open resource coalitions" we recognized the limitations of the traditional model that a system is either working or broken (Figure 3a), because many systems can run in a degraded mode (Figure 3b) [Raz and Shaw 2000 SuffCor].

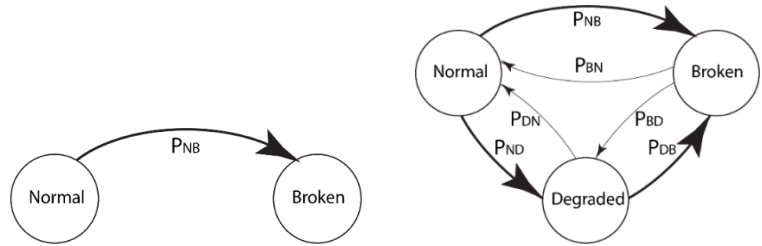

Figure 3.   State transition models for failure. (a) Failure model without repair. (b) State transition model with degraded service and (small) possibility of repair. Transitions are labeled with transition probabilities; recovery transitions have lower probabilities than failure transitions. [Raz and Shaw 2000 SuffCor]

However, this traditional state transition model is too simplistic. First, it requires precise distinctions between states, and degradation can be gradual. Second, real systems can tolerate different degrees of degraded performance, depending on the operating context. In other words, degraded service should be treated as an application-specific continuum, not a sharp state transition. Conceptually, this resembles an island in which ideal performance is at the peak of the island and performance degrades toward water level, with anything underwater corresponding to failure (Figure 4).

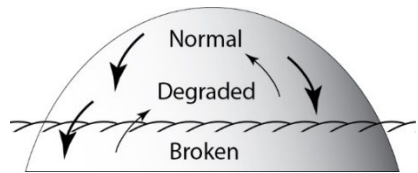

Figure 4.   Degradation and failure in real systems [Raz and Shaw 2000 SuffCor]

The **REFRAMING** of Figure 4 allowed us to think about continuous self-healing, for which we **IMPORTED** the idea of *Homeostasis* from biological systems as described in Example 9.

Example 5.   ***Vernacular Programmers***

Most people who create software are not trained software professionals, but the software research community focuses on supporting the latter. The ad hoc treatment of the other developers, who are largely professionals in some other field, creates **DISSONANCE** between the tools produced to support software professionals and the concepts and tools that would serve them well.

Long before there was a World-Wide Web, non-programmers dominated computer use. They needed to control their own computations, though not by writing programs in the languages of the day, which were designed for systems programming. I argued that their needs are not satisfied by traditional programming languages, and they would be better served if the computing industry created better spreadsheets, symbolic math packages, and other application-specific tools [Shaw 1989 NextPL].

A few years later Barry Boehm predicted that there would be 55 million end user programmers by 2005 [Boehm et al 2005]. In 2005, my student Chris Scaffidi improved this estimate in "Estimating the numbers of end users and end user programmers", suggesting 80 million people using computers at work in 2005 and 90 million by 2012—while there would be fewer than 3 million professional programmers in those years [Scaffidi et al 2005 Number]. He went on to survey information workers about the ways they used spreadsheets, databases, and browsers (there was a World Wide Web by then!). He identified three classes of programming-like activity among these users: recording and using macros, creating database tables linking information with keys, and writing scripts. All respondents were familiar with imperative programming features such as loops, but they did not use them very much [Scaffidi et al 2006 Features].

These users don't simply need to write programs, they should also be concerned with requirements, validation, reliability, maintenance, reuse, and integration in the workflow. However, their engagement with these issues has been unplanned, implicit, reactive, and opportunistic rather than systematic and disciplined. The community of End User Software Engineering researchers addressed this by developing tools and techniques, much of it for spreadsheet programming, to help end users develop better software, more effectively [Ko et al 2011 EUSE]. For example, Scaffidi discovered through contextual inquiry that when administrative computer users had to move data between applications, they often had to re-type the data instead of using cut-and-paste because the number formats were different in the two applications. He developed topes, which could infer conversions between string representations of a data type and allow users to refine, customize, and reuse the definitions [Scaffidi et al 2008 Topes].

The label "end user programmer" has always been awkward and slightly dismissive. When I addressed the myth that programs are written by highly trained professionals in "Myths and mythconceptions: what does it mean to be a programming language, anyhow?" I introduced the term "vernacular programmer" to replace "end user programmer" and discussed the ways that their background, needs, and development styles deserve support from the software community [Shaw 2021 Myth].

### *Pattern 4.  CLASSIFY: Find underlying structure*

> "Gallia est omnis divisa in partes tres"
>
> — Julius Caesar
>
> "Algol is divided into three parts: Subscan, Phase I, and Phase I"
>
> — informal description of CMU Algol compiler, ca 1965-70
>
> "Keep Parks Clean Or Fires Get Started"
>
> — mnemonic for the levels of the Linnaean taxonomy of plants and animals
> (Kingdom, Phylum, Class, Order, Family, Genus, Species)

### *Context*

When complexity challenges our ability to understand a system, we seek ways to bring order. One way to do this is by identifying the important structure—properties, design decisions, constraints--in the system and using it to organize the information in categories that reflect the structure.

Geoffrey Bowker and Susan Star [Bowker and Star 1999] report on how information classification systems shape our views on the systems. They identify two principal ways to develop classifications systems. Some are developed top-down from principled models of the systems; others are developed bottom-up by grouping examples that seem to be related.

We see both kinds in computer science. The Complexity Zoo [Aaronson 2024] is a principled organization of the relations among computational complexity classes, and the card-sorting style of grounded theory is a bottom-up empirical method for grouping and labeling information.

When I worked on the "Software and its Engineering" section of the 2012 ACM Computing Classification System [ACM 2012], I began top-down by separating the topics into categories that correspond to the software itself, to the notations, tools, and scaffolding used to create software, and to the human activities associated with software development. In contrast, the ***Boxology* CLASSIFICATION** of Example 17 arose from identifying similarities and differences among the examples. Note also that Figure 1 and Figure 2 have strong elements of **CLASSIFICATION**.

### *Problem*

*The complexity of the information in a situation, especially the dependencies among elements, is too great to manage informally.*

The information may be partial, or overlapping, or incomplete, or implicit, or represented in different ways. It may come from different stakeholders with different models of the situation. Often it is useful to identify the main themes of the situation by grouping similar information and finding good summary abstractions. If the situation lends itself to an overall model, the model elements may suggest a structure.

### *Solution*

Reorganize the domain description in a structured way that emphasizes the similarities and differences of interest to the current problem.

Many notations and representations are available; what's important is finding a point of view that can apply across the problem.

Bowker and Star give three idealized principles for classification systems:

- There are consistent, unique classificatory principles in operation
- The categories are mutually exclusive
- The systems is complete

They observe that they have never seen a real-world classification that fully meets these requirements and doubt that any practical system ever could. Indeed, practical systems almost all include multiple classification principles, overlapping categories, and a catch-all category for "here are the things that don't fit in an orderly way",

### *Related Patterns*

**Classify** is a specialization of **Reframe**, with emphasis on systematic organization of information about the system.

### *Examples*

Example 6.  ***Software Engineering Research Types***

The late 1990s saw numerous critiques of the lack of rigor in experimental software engineering, some rooted in comparison to physics. This created a strong emphasis on quantitative results, especially quantitative empirical results. However, the pendulum swung far in the direction of favoring quantitative empirical research, which diminished the status of other reasonable paradigms for software research. The field's lack of explicit models for good research set the stage for **Dissonance** about the acceptability of other paradigms.

To assess the spectrum of research strategies that were being practiced and accepted, I read all the abstracts submitted to ICSE 2002 and identified the types of research that were submitted to and accepted by the conference [Shaw 2003 Writing]. After reading all the abstracts (about 300 of them), I settled on comparing them based on the type of research question they were studying, the type of result they produced, and the kind of validation they performed. This **classification** allowed identification of successful research paradigms beyond quantitative empirical studies.

I identified three principal axes of variation in the submissions: the types of research question the papers addressed, the types of results the papers reported, and the kinds of validation carried out.

- The types of research questions addressed development methods, analysis or evaluation methods, design or evaluation of a particular system, generalization, and feasibility study. The acceptance rates for the first three types were notably higher than the acceptance rates for the last two.
- The types of results reported included new procedures or techniques, tools or notations, empirical models, qualitative models, analytic models, solutions to specific problems, and narrative observations. The acceptance rates for the first three were higher than the acceptance rates for the rest; tools were often supporting results.
- The types of validation reported included analysis, experience in use, examples, systematic evaluation, and persuasion; many of the abstracts did not describe validation. The acceptance rates for the first three were higher than the acceptance rates for the others, and papers that relied on persuasion or whose abstracts did not mention validation had very low acceptance rates.

This study only analyzed the abstracts of a single conference, but it has been widely used as a guide for writing up software engineering research. The study was replicated for ICSE 2015, finding that the analysis largely still held, with the addition of research on mining software repositories [Thiesen et al 2018].

Example 7.  ***Design Spaces***

The design alternatives for many problems are rich and open-ended. Nigel Cross observed [Cross 2007] that designers tend to evaluate a broad range of possibilities and refine them selectively, but engineers tend to design depth-first, pursuing the first option that comes to mind and often neglecting alternatives that might yield better results. ***Design Spaces*** help to resolve this **Dissonance** by capturing the design alternatives and their interactions in a structured representation that encourages exploration. They can be used to document design alternatives for a domain, to compare designs, to recommend designs, and in many other ways.

In 1990 Tom Lane interviewed the designers of six user interface systems and identified the functional dimensions (the capabilities of the systems) and the structural dimensions (the implementation choices) of the systems [90LaneDsgnSp]. He then **Classified** them as detailed design spaces and developed a set of design rules that mapped from the desired functional properties to recommendations for implementation. He automated the mapping and supported it with three dozen narrative rules for use by humans. Figure 5 shows the high-level structure of his design space.

| Functional Dimensions | Structural dimensions |
|---|---|
| 1. External requirements | 1. Division of functions & knowledge |
|    a. Application characteristics |    a. Application interface |
|    b. User needs |    b. Device interface |
|    c. I/O devices | 2. Representation issues |
|    d. Computer sys environment |    a. Means of user interface definition |
| 2. Basic interactive behavior |    b. Representation of application info |
|    a. Interface class |    c. Data reps for communication |
|    b. Flexibility of interaction sequencing |    d. Representation of interface state |
| 3. Practical considerations | 3. Control flow, comm, synch issues |
|    a. Portability of applications |    a. Control flow |
|    b. Adaptability of UI system |    b. Communication mechanisms |
|  |    c. Synchronization issues |

Figure 5.  Functional and structural design spaces for user interface structures [90LaneDsgnSp]

The Software Designers in Action workshop provided videos of professional designers sketching a software design for a traffic simulation system to a number of researchers, who responded with individual analyses of the videos [van der Hoek and Petre 2013]. In one of these, I defined a design space to compare the design decisions made by the three teams, the choices implied by the prompt, and the decisions evident in a commercial product [Shaw 2012 design, Shaw 2013 design]. The three teams made very different decisions about the high-level organization of the system, the models of roads and intersections, and even details about how traffic lights would be handled. The representation of this design space is in Figure 6. Each major group of decisions is represented as a tree with two kinds of branches: choice and substructure. Substructure branches (not tagged) group independent design decisions; choice branches, flagged with "##", provide alternatives. In some cases, the decision is a numeric value, and the choices are implicit.

```
Road System                                          Traffic Signals
| High-level organization                            | Place in hierarchy
| | ## Intersections                       AD        | | ## Belong to roads                          AD
| | ## Roads                                         | | ## Belong to intersections                  IN
| | ## Network                             AD  IN    | | ## Belong to approaches,
| Intersections                                      | |    which connect roads to ints             MB
| | ## Collection of signals               IN        | Safety
| | ## Signals and sensors in approaches   MB        | | ## Independent lights with safety checks
| | ## Have roads (with lights and cars)   AD        | | | ##Controller checks dynamically        AD  IN
| Roads                                              | | | ## UI checks at definition time         MB
| | Lanes                                            | | ## One set per intersection, selected from safe set
| | | ## No lanes                                    | Relations among intersections
| | | ## Lanes, with signal per lane      AD  IN    | | ## Independent                              AD
| | Throughput                                       | | ##Synchronized                             IN  MB
| | | Capacity                             AD        | Setting timing
| | | Latency                              IN  MB    | | ## System sets timing                       AD  IN  MB
| Connection of roads to intersections               | | ##Students set timing                      MB
| | ## Intersections have queues (roads)   AD        | Sensors
| | ## Lights and sensors in approaches    MB        | | ## Immediately advance on arrival          IN
| | ## Unspecified or unclear              IN        | | ##Wait to synchronize
| | ## Simulator handles interaction
```

Figure 6. Part of the comparison of several designs for the traffic signal simulator, showing the decisions implied by the task statement (boxed text), made by the three teams (AD, IN, and MB), and made by a commercial product (highlighted) [Shaw 2012 design]

More recently we collected a sample of design spaces, both from the software literature and from the design literature more broadly, to see how they are used in practice [Shaw and Petre 2024 Space]. We found two major themes: some emphasized the principal design decisions and alternative design choices, relegating interactions among the choices to the back burner; others emphasized integrative exploration, keeping interactions front-of-mind. In both cases the design spaces served to systematize understanding of the alternatives, to explore alternatives during design, to support orderly evolution of their understanding, and to capture prior art.

### Pattern 5. IMPORT: Adapt ideas from other fields

"Chance favors the prepared mind."

— Louis Pasteur 1854

"Do you know a related problem? ... Could you use it?
Could you use its result? Could you use its method?"

— George Polya, How to Solve It 1945

### Context

George Polya, in *How to Solve It* [Polya 1945], provided a list of questions to help the reader improve problem solving skills. They included "do you know a related problem? ... Could you use its result? ... its method?" He was, of course, thinking about problems in Euclidian geometry, but the advice clearly applies to looking for inspiration across fields. My father gave me my copy of this book when I was in high school, and everyone should receive such a gift.

Problems have structure, just like systems do. Just as with systems, there's no reason to believe that each field has its own unique set of problems, though most fields have a sec of normal paradigms for dealing with normal problems. Recognizing that a software problem resembles a problem in some other field allows us to ask whether that field's solution suggests an approach for our own, understanding that some re-interpretation may be required.

Expert designers freely import ideas from other projects, other kinds of problems in the same field, and other fields. Among the 66 "ways experts think" identified by Marian Petre and André van der Hoek, almost 10% involve borrowing ideas from previously solved problems, from other artifacts, from opportunistic observation, from analogies. [Petre and van der Hoek 2016]. For example, "Experts look around" (#14), "Experts take inspiration from wherever they can." (#15), "Experts use analogy" (#16), "Experts reshape the problem space" (#20), "Experts explore different perspectives" (#46), "Experts re-assess the landscape" (#61).

It's not reasonable, certainly, to expect that a result from another field will apply directly in software, but it can provide a fresh point of view on the problem, or a technique that could be adapted, or an analogy that inspires a new approach. When borrowing in this way, it's very useful to think about the structure of a problem, not its manifestation in a particular field or subfield. Focusing on the relations among abstractions somewhat removed from implementation details may help you see analogs to your own problem.

We say that you can characterize a field by the types of problems it works on and the kinds of solutions it favors, but that's a sweeping generalization. It's unreasonable to expect that either the problem characterization or the solution will transfer unaltered to software, but this route can often provide ideas and inspiration.[3]

### *Problem*

*After* **Dissonance** *revealed an opportunity, you need a response, but nothing in your current practice springs to mind.*

Think about the structure of the situation that created the dissonance. Do you know of a situation in another field that has similar structure? A similar objective? Similar inputs and outputs? Does it remind you of anything you've seen? Does the description of the dissonance have phrasing that suggests a relation? For example, "cost-effectiveness" could evoke economics,

### *Solution*

Read voraciously. Steal shamelessly. Watch interesting engineering and problem-solving activities outside your own area. Try to understand the conceptual structure of the problems they're solving, not just the implementation of the solution or the labeling of details. Pay attention to the affordances of things in other areas, not just the explicit applications.

When you're working on a problem ask Polya's questions: "Do you know a related problem? … Could you use it? Could you use its result? Could you use its method?" What you find won't exactly match your problem, but see how it can adapt to your context.

### *Related Patterns*

**Importing** often arises in support of **Reframing**.

---

[3] The mapping doesn't even have to be correct: the 19th century process for producing alkali that established industrial processes in England is said to have been based on a flawed analogy to the smelting of metals.

*Examples*

Example 8.  ***Aggregating Research Results***

Research results presented in conference papers do not usually have an easy, direct path to industrial application. They often don't share the same assumptions or context, and they are developed and validated with different methods; in other words, they often do not interoperate well. It's hard to integrate them into larger, more useful results because the incompatibilities must be resolved, and the incentives for doing this are wanting. This **Dissonance** often presents an impediment to moving research results into practice.

Systematic literature reviews offer hope for deriving consensus results, but they struggle to balance the strength of evidence in the results they're aggregating. Even if the synthesized results fall short of a desirable level of confidence, they are often "better than nothing" for practitioners.

Drawing on earlier intuitions about handling evidence with different levels of confidence we developed in work on a ***Calculus of Confidence***, we proposed to map evidence, even weak evidence, to recommendations with techniques that **Import** ideas from Evidence Based Medicine. We recognize several levels of evidence from randomized controlled trials to clinical reports, laboratory studies and case reports to give guidance for translating sets of studies at various levels to strength of recommendation. In "Bridging the gap", we identified the hierarchy of software engineering evidence shown in Figure 7 and recommended further work on how to establish strength of recommendations [Le Goues 2018 Bridge].

| Type of study | | Level | Evidence for Software Engineering |
|---|---|---|---|
| Secondary or filtered studies | | 0 | Systematic reviews with recommendations for practice; meta-analyses |
| Primary Studies | Systematic evidence | 1 | Formal/analytic result with rigorous derivation and proof |
| | | 2 | Quantitative empirical study with careful experimental design and good statistical control |
| | Observational evidence | 3 | Observational result supported by sound qualitative methods, including well-designed case studies |
| | | 4 | Surveys with good sampling and good design; field studies; data mining; crowdsourced evaluations |
| | | 5 | Experience from multiple projects, with analysis and cross-project comparison; Tool, prototype, notation, dataset or other artifact (that has been certified as usable by others) |
| | | 6 | Experience from a single project: objective review of specific project; lessons learned; solution to a specific problem, tested and validated in the context of that problem; in-depth experience report; notation, dataset or unvalidated artifact |
| No design | | 7 | Anecdotes on practice, rule of thumb; validation by toy example, proposed method with careful reasoning, introduction of new concept |
| | | 8 | Position paper, op/ed, etc based principally on expert opinion |

Figure 7.  Hierarchy of evidence for software engineering research [Le Goues 2018 Bridge].

Example 9.  ***Homeostasis***

Adopting the continuous view of degraded service in Figure 4 sets the stage for adding healing mechanisms to systems. In the good-enough realm of Figure 1 it may be harder or more expensive to prevent failure than to notice and repair it. At its simplest this leads to the state transition system of Figure 8a, in which reverse transitions from less-desirable states to more-desirable states are added. This view, however, suffers from the same drawbacks (and **Dissonance**) as the state transition view of Figure 3. To address this, we **Import** the biological concept of ***Homeostasis***,

> Homeostasis is the propensity of a system to automatically resist changes from its normal, or desired, or equilibrium state when the external environment exerts forces to drive it from that state. Software homeostasis as a software system property refers to the capacity for the system to maintain its normal operating state, or the best available approximation to that state, as a result of its normal operation. This operation should both maintain good normal operation and implicitly repair abnormalities, or deviations from expected behavior. [Shaw 2002 Heal]

Figure 8b suggests how incorporating incremental repair as part of the ordinary operation of the system can help to drive the system from broken or degraded performance back to normal operation. Indeed, an ongoing investment in background processes that improve system performance may avoid degradation. Background garbage collection is an example.

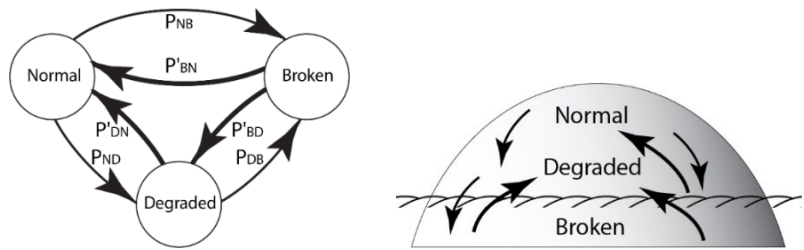

Figure 8.   Self repair. (a) State mode of failure with repair. (b) Degradation, failure, and repair with homeostasis [Raz and Shaw 2000 SuffCor]

Example 10.   ***Design Obligations for Control***

From ***control theory***, we IMPORT the idea of ***feedback loops*** in software systems. In the mid-1970s I was consulting with a chemical process control company, as was Karl Åström, who had just published his now-classic text on control theory [Åström 1984]. This primed me to later bring the ideas, though not the fine details, to software. In control theory, feedback loops are constrained to simple relations, but the reward is complete analysis and guarantees on performance. In software the control relations are richer, so the complete analysis doesn't carry over, but the structure of feedback control reveals design obligations that it's valuable to make explicit.

I first explored this idea as an architectural pattern in "Beyond objects: a software design paradigm based on process control" [Shaw 1995 Control], when I was thinking about the DISSONANCE between claims that objects would solve all problems and practical software. In this example I compared control and object-oriented approaches to 1990s-era automotive cruise control and showed how the control view raised design questions that were ignored in the object view; these included accuracy of the model's value for speed, whether the system could reduce as well as increase speed, and how state and event inputs interact. Automotive cruise control serves software engineering as a "model system"[4], and in "Comparing architectural design styles" [Shaw 1995 Style] I compared eleven cruise control designs, including examples of object-oriented, state-based, feedback control, and real-time architectures. The comparison showed that the choice of architecture shaped the issues the designer dealt with as well as the ability to check the design.

---

[4] It is common for a discipline, especially one that is just getting its wits about itself, to adopt some shared, well-defined problems for teaching and study. Known as *model systems* or *type problems*, they provide a way to compare methods and results, develop new techniques on standard examples, and set a minimum standard of capability for new participants. A reasonable approach to some of these problems becomes the threshold to get serious consideration of a new technique. Type problems also provide a pre-debugged source of educational exercises. [Shaw et al 1995 ModelProb]. "Stack" served this role for abstract data types, as did "dining philosophers" for synchronization. The house mouse *Mus musculus* and fruit fly *Drosophila melanogaster* serve as type systems in genetics [White 2016].

***Design obligations*** are design-level analogs of proof obligations. For example, proving loop invariants in a programming languages requires an inductive proof. Also, Tony Hoare documented the proof obligations for the representations of abstract data types [Hoare 1972]. Similarly, some architectures rely on specific relations among components that impose design obligations to establish and preserve these critical relations. For example, database systems rely on the ACID properties (atomicity, consistency, isolation, and durability), and true pipe-and-filter systems assume no out-of-band communication between the filters. In feedback-based adaptive control, a set of obligations arise from the relations that must be assured among elements of the feedback process, as described in Figure 9.

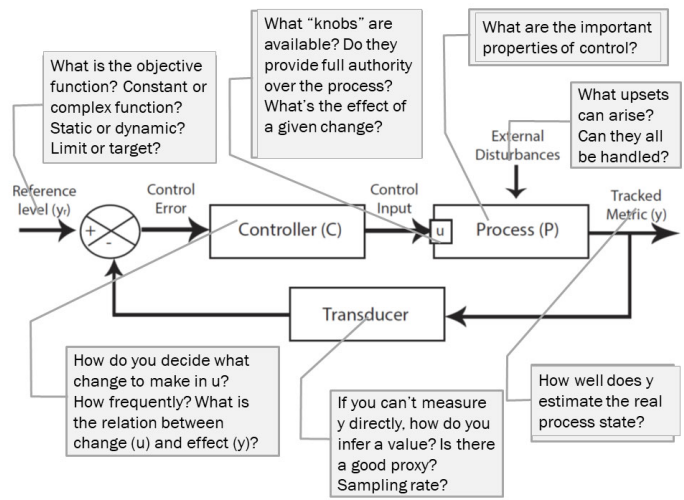

Figure 9.   Design obligations for control [Shaw 2016 Control]

I returned to control in the context of self-adaptive systems with a plea for "Visibility of control in adaptive systems" [Müller et al 2008 Control]. We were frustrated by the **Dissonance** between ideas from control theory and the lack of attention to control properties in the commonly used MAPE architecture for adaptive systems. We called for designers be explicit about how the adaptive systems actually achieve control. For example, the designer of an adaptive system that is supposed to maintain some parameter at a reference value should show that the system is controllable (the reference value is achievable), stable (control actions do move the system to the reference value), and robust (it's stable despite external disturbances).

At a later Dagstuhl on self-adaptive systems [Dagstuhl 2017], we explored in depth the application of control theory to self-adaptive systems. In "What can control theory teach us about assurances in self-adaptive software systems?" [Litoiu et al 2017 CtlThy] we identified the limitations of classical control theory for large-scale adaptive systems, showed how aspects of classical control could nonetheless guide design, and worked through several case studies. These ideas are perhaps more accessible in my talk "What can control theory teach us about designing cyber-physical systems" [Shaw 2016 Control], which examines the components of a feedback loop and identifies design obligations—properties that the designer must address—as shown in Figure 9. These are essential questions that the usual software development methods and tools, including UML, do not address.

### *Pattern 6.  SATIRIZE: Ridicule absurdity*

> "Anything worth doing is worth doing to excess"
>
> — Edwin Land

**Context**

Humor, especially satire, can be a safe way to make statements that might otherwise be unacceptable. This has been true since court jesters not only provided entertainment, but could also exercise "jester's privilege"—the right to mock freely without being punished. Taking an idea to extremes, vividly pointing out contradictions between principle and policy, or ridiculing it in other way have the potential to push people past the point of tolerating an annoyance to recognizing a need deal with a problem. Even gentle joshing has its place.

Policies, pronouncements, mandates, requirements, or standards are sometimes promulgated in ways that we can't disobey, but compliance leads to absurd outcomes. Ridiculing them provides a way to vent, and sometimes it draws attention to the absurdity. I can't point to any instance in which it made an immediate difference, but it can contribute to community pushback. In any case, satire does entertain people and calls out the problem in a memorable way: My 1971 curse on the IBM 360 [Shaw 1971 Curse], which ranted about the machine's poor usability, is still floating around the Internet.

Satire can also be used to test assumptions. Driving assumptions to their limits can expose unexpected emergent behavior in a system, and it's better to find these with humor than in real-life crises.

**Problem**

*A policy, pronouncement, requirement, process, measure of merit, or standard is promulgated, without consultation or evidence and resources for satisfying it. The policy cannot be rejected—for political, rather than substantive reasons—but it leads to absurd outcomes.*

It sometimes takes concerted pushback or widespread criticism to reverse or repair actions like these. Humor can help to gather attention and support.

**Solution**

Make fun of the absurdity, especially in informal channels. Make it look as ridiculous as possible, for example by exaggeration or by malicious compliance. People can accept a certain amount of conceptual inconvenience—or **DISSONANCE** between principle and practice. However, there's often a breaking point, and satire can push past the breaking point.

So, write it up for your favorite humor venue. *SIGBOVIK* is CMU's conference patterned after the *Journal of Irreproducible Results*. It's a great outlet for whimsy and satire. I turn to this venue when I want to make fun of some absurd policy or requirement. Most often, I publish under some *nom de farce*. More ambitiously, try for *McSweeny's*, or *The Onion*, or *Journal of Irreproducible Results*, if those outlets still exist.

Other uses of humor, such as whimsy, can increase engagement with ideas.

**Related Patterns**

**REFRAMING** can sometimes redirect attention to important aspects of the underlying problem. **IMPORT** may suggest an analogy that lets you parody your problem in an improbable way.

### *Examples*

Example 11. **Push the envelope**

Sometimes the way to make fun of administrative overkill is to push it down its slippery slope on a toboggan and see where it lands.

A doctoral program at a Certain Major University reviews the progress of all students every semester. On one occasion the excuses for lack of progress across the students were many and varied, much more so than usual. They included "changed research areas", "served on admissions committee", "submitted material too late to review", "changed advisors", and "my cat died". We reacted by attributing all the excuses, plus some from previous years, to a single hypothetical "professional student" and publishing a short report showing a 12-year path to still not having a thesis topic. The resulting report, "The professional student's strategy for perpetual funding" appeared as a technical report [Bovik 1993 Perpetual]. It had staying power, being recognized as the "Most Influential Paper from $2^{2^{2^{2^0}}}$ years ago" at the 2009 SIGBOVIK conference [Bovik 2009 MIP].

A Certain Major University also used to send an 8-page memo to faculty each year with a list of religious holidays for the current year and a request to minimize scheduling assignments on these dates. The memo included Christian, Jewish, Hindu, Islamic, Buddhist, Shinto, Jain, Sikh, Baha'i, Zoroastrian, Wiccan, and other observance dates. The **DISSONANCE** between the long list of conflicts and the pedagogical need to make at least some assignments led inevitably to the question, "Is there a set of observances that completely cover the academic calendar, thereby precluding all due dates?"

The dates in the memo covered well over half of the class days. The question of full coverage was an interesting exercise, because the rules for the dates of annual observances follow different patterns, so a covering set of holidays for one year is unlikely to be a covering set for another year. Common holidays follow cyclic patterns, though different ones (explicit date vs day in the $n^{th}$ week of the month), and they are affected by leap years. More challenging are the complex interactions of calendars with different cultural bases: (Gregorian, Julian, lunar calendars) and events with late-binding dates determined by celestial observations each year. [Fidget and Nowhey 2011 Holiday].

We concluded that there are enough recognized holidays to provide complete coverage, perhaps with a bit of research. In other words, the university memo moved the needle from observing the holidays of one religion to observing the holidays of several religions (a worthy goal, but **DISSONANT** with the educational mission, and we pushed the needle all the way over to full coverage. In other words, we moved the needle:

| Gimme that old time religion, | Gimme that old time religion, |
|---|---|
| Gimme that old time religion, | Gimme that new-age religion, |
| Gimme that old time religion, ⟶ | Gimme that weird off-beat religion, |
| It's good enough for me. | There'll be NO deadlines for me. |
| Adapted from [Gospel 1873] | [Fidget and Nowhey 2011 Holiday] |

The university in question no longer distributes the 8-page memo.

Example 12. **Thinking, not counting**

Between the tXitter-driven passion for reducing complex ideas to simplistic statements and the current fashion for numerical evidence, we are plagued with charts and numbers that offer the illusion of precision with little assurance (and hence considerable **DISSONANCE**) that the numbers are actually accurate or that they actually support the conclusions drawn from them. Examples close to home include university and computer science department rankings and the academic passion for bibliometrics. Both of these are ad hoc quantitative algorithms that have only tenuous connections to quality and therefore lend themselves to ridicule. The CSrankings.org web site claims to solve the subjectivity problem in rankings by having an algorithm that is both objective and transparent. However, I used **Satire** to show the deficiency of that as the sole criterion by providing an algorithm EvenbetterCSRankings.org that is objective, transparent, and superior against other criteria but whose output is ridiculous [Diogenes 2021 Rankings].

Bibliometrics such as citation counts and the h-index are similarly poor proxies for quality (people mostly look at Google Scholar, because it vacuums up more material than others; it tends to run about 4 times as high as the ACM citations counts). The substantive issue is that the counts do not take account of the context of the citation, and the numbers are not comparable across fields, especially fields with very different publication expectations and different numbers of researchers.

Complaints about bibliometrics tend to focus on these and the way they are abused in evaluating individuals. We took on a somewhat more esoteric ego metric, the Erdös number and its extension, the Erdös-Bacon-Sabbath number, which measure collaboration distance to Paul Erdös (scientific publications), Kevin Bacon (movie credits), and Black Sabbath (performance). A collaborative group including members with low indexes produced a **Satirical** documentary about a performance and wrote a paper to demonstrate how easily these numbers can be manipulated (and my Erdös-Bacon-Sabbath number is now a remarkably low 7) [Klawe et al 2020 EBS]. The point, of course, is to shame people into actually thinking about their evaluations rather than blindly slinging numbers.

Example 13.  ***Whimsy***

This is perhaps the gentlest form of **Satire**, playful or fanciful humor. I used I used it in context to add a distinctive note to papers I published at ICSE and related workshops in 2000, in Limerick Ireland. Of course, this conference was in Limerick, so it must have limericks. For each paper I provided a synopsis, in the form of a limerick.

A coder of software was glad
For the "engineer" title he had.
　　But he'd learned just to hack,
　　Not to think, choose, or track.
So the client's big startup went bad :-(

This one tied for the "best limerick" award at ICSE 2000. It summarizes "Software engineering education: a roadmap". Alas, it appeared in the talk but not the published paper.
[Shaw 2000 Ed]

Proving system correctness is tough.
It can fail, or succeed, or just bluff.
　　When the parts come and go
　　You may never quite know …
Can you tell when just "good" is enough?

"Sufficient correctness and homeostasis in open resource coalitions: how much can you trust your software system?"
[Shaw 2000 Homeostasis]

My model for choosing investments
Expects to get ratio-scale measurements
　　But for software design
　　Even ordinal is fine
Can the model give reasonable guesstimates?

"When good models meet bad data: applying quantitative economic models to qualitative engineering judgments"
[Butler et al 2000 GoodBad]

## A Sequence of AHA! Pattern Uses

### 4  *SOFTWARE ARCHITECTURE*

Among the technologies that promised universal solutions, the flourishing of object-oriented programming in the 1970s and 80s brought exaggerated claims that objects would solve all our problems. It's possible that all software can be somehow cast as objects, but that perspective often isn't very helpful, as we found when we tried to define compilers, then text editors, as objects: We could write the definitions, but the essence of compiler-ness and text-editor-ness were lost.

Building on a long tradition of recognizing "hip-pocket algorithms that everyone should know", in the 1980s we stepped back from objects and started thinking about "hip-pocket program skeletons that everyone should know".

At that time, programmers had informal vocabulary for describing program organizations. We could fill blackboards and bar napkins with freehand box-and-line figures while using words like filter, object, process, pipeline, transaction, and the like. We still implemented these ideas with procedure calls between compilation units, or perhaps inter-process calls. However, practice was at best ad hoc.

The **DISSONANCE** between the abstract ideas in what was essentially folklore and the formal proclamations about objects and filters led me to **REFRAME** the description of software system structures. Initially I studied several ***Alternatives to Objects***, studying cases where the object point of view was unhelpful and practice was ad hoc. Subsequently I returned to an early example of ***Alternatives to Objects***, control for self-adaptive systems, by **IMPORTING** feedback ideas from control theory to lay out ***Design Obligations for Control*** architectures (Example 10 of the **IMPORT** pattern).

Recognizing that the concepts were similar to the hardware architecture level of computer system design, which addresses the relation of components such as processors, memories, and switches, I **IMPORTED** the ***Computer Architecture Concepts*** of components and connectors [Bell and Newell 1970].

The ensuing discussions produced a chain of insights that identified popular abstractions as a set of ***Software Architecture Styles*** and **REFRAMING** the organization and analysis of software system structures through that lens. After those initial styles became established, we **CLASSIFIED** them (and their box-and-line diagrams) in a ***Design Space*** for architectural styles called ***Boxology***.

To support the use of the abstractions central to software architecture styles, I **REFRAMED** notations for these software structures as ***Languages for Software Architecture***, which identified the representations and language concepts need to step up from linking modules with procedure calls to directly invoking the abstractions about components and connectors that underlie architectural styles.

The formal programming languages community objected to the (lack of) ***Formality of Software Architecture*** on the grounds that the languages were not minimal and not sound. Indeed, the architectural abstractions could be coded in conventional languages, but that obscured the abstractions. Indeed, these languages are not sound; they describe compositions of components that are neither fully specified nor verified, and soundness over unsound primitives would not contribute much. Therefore it was appropriate to **SATISFICE** and reject assumption that programming languages must be sound and fully general in order to be useful.

The examples here describe some of the creative steps in the research. These and other contributions come together in "Software architecture: perspectives on an emerging discipline" [Shaw and Garlan 1996 SWArch].

Example 14.  ***Alternatives to objects***

The 1980s enthusiasm for object-oriented design led to sweeping claims about how object orientation was always the "best way" to organize software. However, it often appeared that the object point of view failed to capture the essential ideas about a system and, depending on the system, defining the system in some other model gave a better sense of the essence of the design. This **Dissonance** led me to explore ways to think abstractly about software systems other than the object view.

For example, a compiler computes a transformation from source code to object code. Thinking of it as a object isn't particularly helpful, because that doesn't provide insight into how the transformation takes place. At the time, the common diagram of a compiler was a linear pipeline of several phases (lexing, parsing, semantics, optimization, code generation, etc) as depicted in Figure 10. These transformations form the essential idea; certainly there can be objects in the implementation, but the main idea, and the main concern of correctness, is the sequence of transformations.

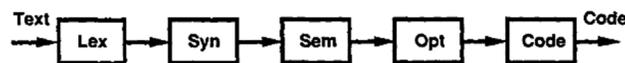

Figure 10. Classical view of a compiler architecture [Shaw 1996 SIS]

As compilation techniques became more sophisticated, the intermediate data structures became more prominent. Improved theory, such as attribute grammars, guided the model away from the pipeline model. The symbol table became an external data structure shared by the phases, and the intermediate representation (the attributed parse tree) also shared data structure that was initialized early in compilation and augmented through the course of compilation. However, the common representation of the compiler remained in the classical pipeline as depicted in Figure 11.

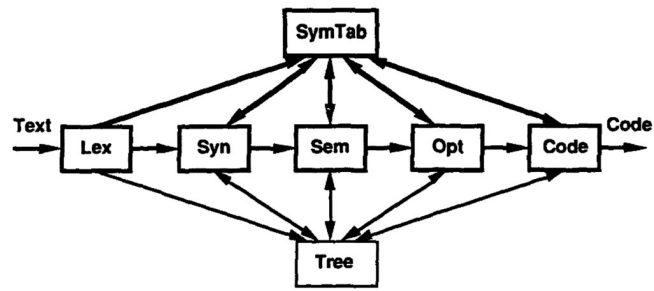

Figure 11. Compiler with symbol table and parse tree shared by phases [Shaw 1996 SIS]

In "Software architecture for shared information systems" I argued that a more appropriate representation would recognize how much the architecture of the compiler had moved away from the original pipeline. In modern compilers the system structure had become a data-centered repository, similar to a database, that was operated on by a series of processes as depicted in Figure 12 [Shaw 1996 SIS]. This **Reframing** would have made it easier to discuss the order in which to apply optimizations, a discussion that was constrained at the time by the notion that there should be a fixed order for the optimizations.

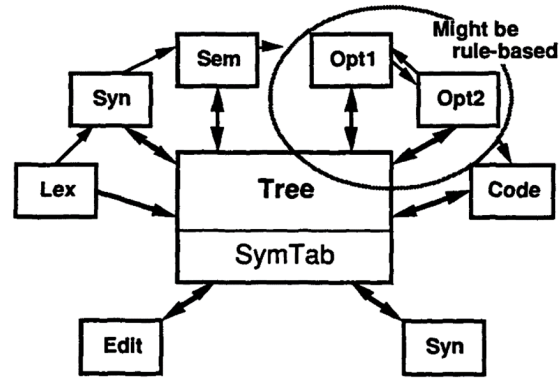

Figure 12. View of a compiler as a data-centered repository [Shaw 1996 SIS]

Moving beyond the unhelpfulness of objects for explaining compilers, I began systematically exploring **Alternatives to objects** with a simple example, the inappropriateness of classical programming models of interactive input-output in "An input-output model for interactive systems" [Shaw 1986 IO]. Here I showed how the then-current model of I/O as simple parsing/generation of text streams broke down in the then-new interactive environments with rich, interactive visualizations of values. I proposed a richer model that **Reframed** the interactions with explicit display state, dynamic update, and type-specific processing. Subsequently my student Tom Lane explored this in more depth in "Studying software architecture through design spaces and rules," by developing and validating a **Design Space** for interactive user interfaces and providing automated design rules for selecting a user interface architecture based on the needs of each specific system (Figure 5 of the **Classify** pattern) [Lane 1990 Design Space].

My next specific model of an **Alternative to Objects** was adaptive systems. For this I **Imported** ideas of control theory. Initially I proposed a qualitative model based on feedback loops using automobile cruise control as an example. Much more recently, I revisited the idea in more detail and identified **Design obligations for control**, which not only explored the imported idea but also established the need for design obligations to accompany architectural styles (Example 10 of the **Import** pattern).

These examples led to the broader question of what other software structures are in common use and what system-level abstractions are useful, the genesis of **Software Architecture**, inspired by **Importing** hardware configuration ideas from **Computer Architecture**

Example 15. **Computer Architecture Concepts**

From **Computer Architecture**, we **Import** the concept that computers are described at several levels of detail, from electrical circuits to the actual components that make up a computer system. Gordon Bell and Allen Newell termed this the processor-memory-switch level, and they defined PMS, a computer architecture description language to describe this level of organization [Bell and Newell 1970]. The language describes computers or networks as linked collections of seven basic component types: **P**rocessor, **M**emory, **S**witch, **K**(c)ontrol, **D**ata, **L**ink, and **T**ransducer; the notation shows the connections between them and annotations to show which variants of the components are used. For example, Figure 13 depicts a LINC-8-388, which was a PDP-8 with LINC processor and 388 display)

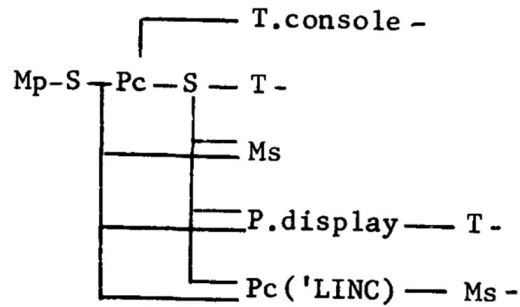

Figure 13. PMS diagram of LINC-8-388 [Bell and Newell 1970]

This view of computer architecture, specifically the distinctions about how hardware components interacted with each other (the cables connections are different, so you can't connect the cables in arbitrary ways!), was the direct inspiration for my view of ***Software Architecture*** as the high-level structure of a software system, composed of components of various distinguished types interacting via connectors of various distinguished types. This connection is explicit in "Elements of a design languages for software architecture" [Shaw 1990 ArchLang]; the connection is clear but less explicit in the original 1988 exposition of the idea, "Toward higher-level abstractions for software systems" [Shaw 1988 Abstr]. We later fleshed out the character of a ***Language for Software Architecture***, making types of components and types of connectors first-class constructs in the language. From those roots arose the **Reframing** of module composition from the classical provides/requires model to a model based on abstractions for components and connectors.

Example 16. ***Software Architecture Styles***

In the 1980s it was common to publish papers about the implementation of software systems. These papers usually included about six column inches about the "architecture" of the system, which included a box-and-line diagram of about two column inches and text describing the system. This text used a rich, but informal, idiomatic vocabulary of abstractions that had evolved over time without standardization. Although the vocabulary and the text that used were imprecise and informal, designers nevertheless communicated with some success. For example:

> "Camelot is based on the client-server model and uses remote procedure calls both locally and remotely to provide communication among applications and servers." [Spector 1987]

> "Abstraction layering and system decomposition provide the appearance of system uniformity to clients, yet allow Helix to accommodate a diversity of autonomous devices. The architecture encourages a client-server model for the structuring of applications." [Fridrich 1985]

Around 1990 I collected a bundle of these papers, extracted the architecture description, and clustered them into groups that shared vocabulary and used similar diagrams. Several common patterns emerged, which I **Reframed** as architectural styles. I described them informally in "Larger scale systems require higher-level abstractions" [Shaw 1988 Abstr] and then more systematically in the first and second PLoP conferences as "Patterns for software architectures" [Shaw 1994 PLoP] and "Some patterns for software architectures" [Shaw 1995 PLoP]. Sample box-and-line diagrams for a few of these styles are in Figure 14.

Significant in these diagrams is the way that the interactions between the boxes are labeled. These are not simply procedure calls, they are abstractions about patterns of interaction: Yes, there are subroutines in Figure 14a, b, and c. because those represents the classical module structure; as abstract data types became subsumed by object systems, the calls of Figure 14b became method invocations, and the calls of Figure 14c are specialized system calls. More significantly, Figure 14d uses implicit invocation, and Figure 14e uses unix-style data flow in ASCII pipelines.

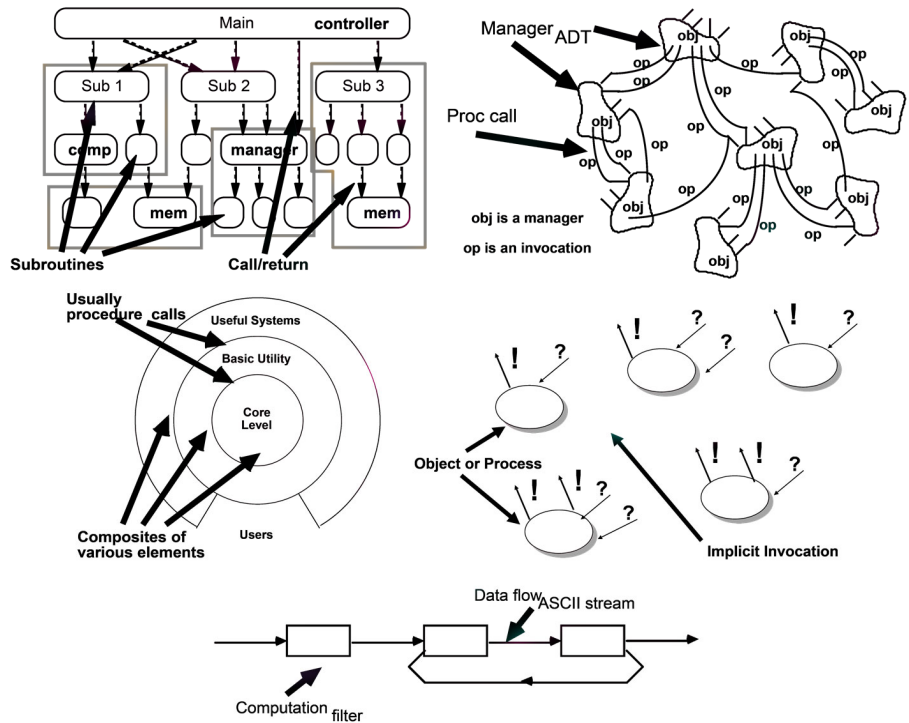

Figure 14. Box-and-line diagrams for architectural styles: (a) main program and subroutines; (b) data abstraction; (c) layers; (d) implicit invocation; (e) pipeline [Shaw 1995 PLoP].

These styles, of course, are not used in isolation or in their pure form. The three case studies of "Heterogeneous design idioms for software architecture" [Shaw 1991 Idiom] apply the styles to interpret three systems that were designed without reference to the styles.

Example 17.  **Boxology Design Space**

Further exploration of the styles led to systematic classification of the architectures represented in box-and-line diagrams and eventually to "A field guide to boxology: Preliminary classification of architectural styles for software systems" [Shaw and Clements 1997 Boxology]. This CLASSIFICATION helps software designers choose an architecture suitable for the problem at hand. Supporting those choices requires careful discrimination among the candidates and guidance on making appropriate design choices.

This is, of course, a **Design Space** as discussed in Example 7. It is a 2-dimensional description with abstract architectural styles as rows, clustered by architectural style. Columns capture details such as the constituent components and connectors and issues related to control, data, and their interaction. The classification also identifies the type of reasoning that commonly applies. Figure 15 shows a snippet covering only data-centered repositories, about 20% of that design space.

In addition to defining the design space to describe the varieties of architectural styles, we began the task of providing design rules of the general form "If your problem has characteristic X, consider architectures with characteristic Y". This follows Lane's guidance for user interface systems, as discussed in Example 7. Some examples of these design rules are [Shaw and Clements 1997 Boxology]

- If your problem involves transformations on continuous streams of data (or on very long streams), consider a pipeline architecture.
  - However, if your problem involves passing rich data representations, avoid pipelines restricted to ASCII.
- If your system involves controlling continuing action, is embedded in a physical system, and is subject to unpredictable external perturbation so that preset algorithms go awry, consider a closed loop control architecture.
- If your task requires a high degree of flexibility/configurability, loose coupling between tasks, and reactive tasks, consider interacting processes.
  - If you have reason not to bind the recipients of signals from their originators, consider an event architecture.
  - If the tasks are of a hierarchical nature, consider a replicated worker or heartbeat style.
  - If the tasks are divided between producers and consumers, consider client/server.

| Style | Constituent parts | | Control issues | | | Data issues | | | | Control/data interaction | | Type of reasoning |
|---|---|---|---|---|---|---|---|---|---|---|---|---|
| | Components | Connectors | Topo-logy | Synch-ronicity | Binding time | Topo-logy | Contin-uity | Mode | Binding time | Isomorphic shapes | Flow directions | |
| **Data-centered repository styles:** Styles dominated by a complex central data store, manipulated by independent computations | | | | | | | | | | | | Data integrity |
| Transactional data-base [Bc90, Sp87] | memory, computations | trans. streams (queries) | star | asynch, opp | w | star | spor lvol | shared, passed | w | possibly | if isomorph-ic, opposite | ACID[5] properties |
| •Client/server | managers, computations | transaction opns with history[3] | star | asynch. | w, c, r | star | spor lvol | passed | w, c, r | yes | opposite | |
| Blackboard [Ni86] | memory, computations | direct access | star | asynch, opp | w | star | spor lvol | shared, mcast | w | no | n/a | convergence |
| Modern compiler [SG96] | memory, computations | procedure call | star | seq | w | star | spor lvol | shared | w | no | n/a | invariants on parse tree |
| **Key to column entries** | | | | | | | | | | | | |
| Topology | hier (hierarchical), arb (arbitrary), star, linear (one-way), fixed (determined by style) | | | | | | | | | | | |
| Synchronicity | seq (sequential, one thread of control), ls/par (lockstep parallel), synch (synchronous), asynch (asynchronous), opp (opportunistic) | | | | | | | | | | | |
| Binding time | w (write-time--that is, in source code), c (compile-time), i (invocation-time), r (run-time) | | | | | | | | | | | |
| Continuity | spor (sporadic), cont (continuous), hvol (high-volume), lvol (low-volume) | | | | | | | | | | | |
| Mode | shared, passed, bdcast (broadcast), mcast (multicast), ci/co (copy-in/copy-out) | | | | | | | | | | | |

Figure 15. Snippet of the Boxology design space, for data-centered repositories [Shaw and Clements 1997 Boxology]

Example 18. ***Languages for Software Architecture***

The programing language view of modules is that they comprise code and some procedures that that can be called to invoke execution of the code on internal data structures. Their interface definitions list the procedures, exported data, and perhaps types and exceptions. The modules have equal status; they are not differentiated by type.

In conventional languages, nothing akin to a type system distinguishes different sorts of modules, for example groups of procedures from objects from filters from data stores. Modules interact largely via procedure calls or specializations like remote procedure calls or method invocations (direct data access is also possible but has been generally frowned upon for half a century [Wulf and Shaw 1973 Global], except in specific controlled ways). Nothing akin to a type system distinguishes different forms of interaction, say data flow, implicit triggering, message passing, and transactions. Those abstractions are all used informally in design, but they are implemented with procedure calls because that's the construct provided by programming languages.

However, progress in programming languages is measured by how we raise the level of abstraction [Shaw 1980 Abstr], and these more abstract constructs are the appropriate ones for software architecture. This **Dissonance** between the abstractions appropriate to design and the available programming languages triggered **Reframing** of the language to the architectural level of abstraction.

Raising the level of abstraction in the programming language requires mapping to the implementation technology, which is procedure calls. In "Procedure calls are the assembly language of software interconnection" I elaborated the rationale for designing with architectural abstractions, especially with *connectors* that abstract the various forms of interaction that are usually implicit [Shaw 1993 Assy].

My particular innovation in architecture description languages was introducing abstract *connector* types in addition to *component* types as part of the architecture description language. In "Toward higher-level abstractions", we called out identifiable types of interactions between components and called them "connectors" [Shaw 1988 Abstr]. These included procedure call, data flow, implicit triggering, message passing, shared data, and instantiation. Consider a traditional box-and-line diagram as in Figure 16, as it might typically have been drawn at the time.

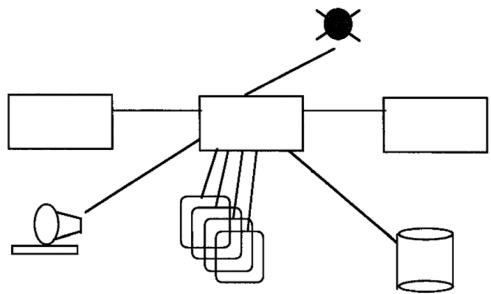

Figure 16. Typical box-and-line depiction of a software architecture [Shaw 1993 Assy]

The shapes of the boxes may suggest their function but the lines simply show that there is some association. Reflection on the hardware example of the PMS language suggested that the relations indicated by those lines are of identifiable types, as indicated by Figure 17a; this allows the box-and-line diagram to be redrawn as in Figure 17b, which is much more informative about the design.

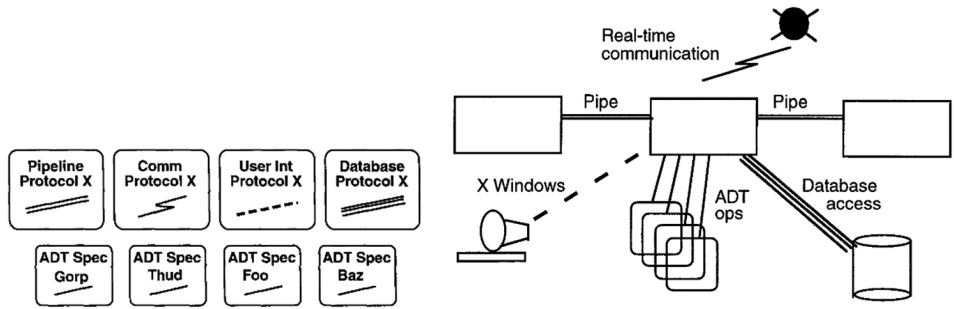

Figure 17. (a) Constellation of protocol specifications required by example.
(b) Revised architecture diagram with discrimination among connections [Shaw 1993 Assy]

Introducing the concept of connectors in the architecture description language raises the abstractions about these interactions to first-class status. It localizes the information about the protocols, and it sets the stage for type-checking of these abstractions.

Example 19.  *Formality of Software Architecture*

The theoretical programming language research community holds that programming languages should be sound and complete, that the formal language definition should be derivable from minimal primitive definitions. However, a language for software architecture addresses the way components are organized (and especially how they interact), independent of whether the components themselves are correct or even completely specified. Further, the components are not all code modules: they include data sets, online data feeds, interactive components, and so on. As for symbolic rigor, much of the information about system-level software is qualitative. There is substantial **Dissonance** between these views.

So I **Satisficed**, taking the position that fitness to task is more important than minimality at the software architecture level of abstraction, so supporting abstractions for the high-level organization of software is precisely the right thing to do. We can still reason at the system level even if the parts are fallible.

The theoretical community objected specifically to the concept of "connector" identified by **Import**. They argued that it's unnecessary because it can be defined from other primitives; the same module concept could be generalized to cover both components and connectors. Indeed, it can. However, at the architecture level, components and connectors serve different roles. Making connectors first-class elements of an architecture description language localizes definitions, provides type-like concepts for important abstractions, and identifies the code fragments that are often distributed through code to achieve the protocol that implements the abstraction.  I stand by this decision.

## ACKNOWLEDGMENTS

This work draws heavily on ideas from engineering and design, especially from the colleagues cited in the general references. The thoughtful critique and discussion from the 2023 PLoP workshop led to major improvements in the paper. Richard Gabriel's extensive shepherding in advance of the workshop shaped the development of the paper, and the members of the workshop group--Indu Alagarsamy, Andrew Black, Marden Neubert, Doug Schmidt, Kurihara Wataru, and Joe Yoder—provided valuable feedback that has improved this final version. Marian Peter has been a thoughtfully critical sounding board throughout. Discussions with my numerous collaborators through the years have provided the back-and-forth from which dissonance appears. This work is supported by the Alan J. Perlis Chair of Computer Science at Carnegie Mellon University.

## REFERENCES

**General references (by author)**

[Aaronson 2024] Scott Aaronson. *Complexity Zoo.* https://complexityzoo.net/Complexity_Zoo (accessed 2/8/2024).

[ACM 2012] Association for Computing Machinery. *The 2012 ACM Computing Classification System.* https://www.acm.org/publications/class-2012 (accessed 2/8/2024)

[Åström 1984] Karl Åström. *Computer Controlled Systems: Theory and Design.* Prentice-Hall, first edition, 1984.

[Bell and Newell 1970] C. Gordon Bell and Allen Newell. The PMS and ISP descriptive systems for computer structures. *AFIPS '70 (Spring): Proceedings of the May 5-7, 1970, spring joint computer conference*, May 1970 Pages 351–374 https://doi.org/10.1145/1476936.1476993

[Boehm et al 1995]. Barry Boehm et al. Cost models for future software life cycle processes: COCOMO 2.0. *Ann Software Eng 1*, 57–94 (1995). doi: 10.1007/BF02249046

[Boehm 1981] Barry W. Boehm. *Software Engineering Economics.* Prentice-Hall 1981

[Bowker and Star 1999] Geoffrey C. Bowker and Susan Leigh Star. *Sorting Things Out: Classification and its consequences.* MIT Press 1999.

[Card et al 1980] Stuart K. Card, Thomas P. Moran, and Allen Newell. 1980. The keystroke-level model for user

performance time with interactive systems. *Comm. ACM* 23, 7 (July 1980), 396–410.

[Cheng 2023] Eugenia Cheng. *The Joy of Abstraction: an exploration of math, category theory, and life.* Cambridge University Press 2023.

[Cross 2007] Nigel Cross. *Designerly Ways of Knowing.* Board of International Research in Design. Birkhäuser Architecture 2007.

[Dagstuhl 2017] R. de Lemos, D. Garlan, C. Ghezzi, H. Giese (eds) *Software Engineering for Self-Adaptive Systems III, Assurances.* Lecture Notes in Computer Science vol 9640, Springer, 2017, pp 90–134.

[Foley 2000] Mary Jo Foley, Sm@rt Reseller. Bugfest! Win2000 has 63,000 'defects'. *ZDNet News*, February 2000. Full article at https://www.zdnet.com/article/bugfest-win2000-has-63000-defects/ (accessed 8/23/2023).

[Fridrich 1985] Marek Fridrich and William Older. Helix: The Architecture of the XMS Distributed File System. *IEEE Software*, vol 2, no 3, May 1985 (pp. 21-29).

[Gospel 1873] Traditional. Old Time Religion. Documented in 1873 list of traditional Gospel songs, possibly earlier. Public domain.

[Hoare 1972] C. A. R. Hoare. Proof of correctness of data representations. *Acta Informatica* 1(4), 271-281 (1972). https://doi.org/10.1007/BF00289507

[Knuth 1969] Donald E. Knuth. *The Art of Computer Programming: Semi-Numerical Algorithms.* Addison-Wesley, 1969.

[Korzybski 1933] Alfred Korzbyski. *Science and Sanity: An Introduction to Non-Aristotelian Systems and General Semantics.* First edition 1933. Fifth edition 1993, Institute of General Semantics.

[Lakatos 1976] Imre Lakatos. *Proofs and Refutations: The Logic of Mathematical Discovery.* Cambridge University Press 1976.

[Polya 1945] George Polya. *How To Solve it.* Princeton University Press 1945.

[Schön 1983] Donald A. Schön. *The Reflective Practitioner: How Professionals Think in Action.* Basic Books 1983.

[Petre and van der Hoek 2016] Marian Petre and André van der Hoek. *Software Design Decoded: 66 Ways Experts Think.* MIT Press 2016.

[Rittel and Webber 1973] Horst W. J. Rittel and Melvin M. Webber. 1973. Dilemmas in a general theory of planning. *Policy Sciences* 4, 155–169 (1973). https://doi.org/10.1007/BF01405730.

[Shapiro and Varian 1998] Carl Shapiro and Hal R. Varian. *Information Rules: A strategic Guide to the Network Economy.* Harvard Business Review Press 1998.

[Simon 1956] Herbert A. Simon. Rational choice and the structure of the environment. *Psych Review* 63(2), 129–138.

[Simon 1979] Herbert A. Simon. Rational decision making in business organizations. *The American Economic Review.* 69 (4): 493–513.JSTOR 1808698.

[Spector 1987] Alfred Z. Spector et al. Camelot: A Distributed Transaction Facility for Mach and the Internet. An Interim Report. *Carnegie Mellon University Computer Science Technical Report*, June 1987.

[Thiesen et al 2018] Christopher Theisen, Marcel Dunaiski, Laurie Williams, and Willem Visser. Software Engineering Research at the International Conference on Software Engineering. 2016. *SIGSOFT Softw. Eng. Notes* 42, 4, January 2018.

[van der Hoek and Petre 2013] André van der Hoek & Marian Petre (eds). *Software Designers in Action: a Human-Centric Look at Design Work.* Chapman and Hall/CRC, 2013.

[White 2016] Benjamin H White. What genetic model organisms offer the study of behavior and neural circuits. J *Neurogenet.* 2016 Jun;30(2):54-61. doi: 10.1080/01677063.2016.1177049

**References for the examples (by year)**

[Shaw 1971 Curse] Mary Shaw. Curse/360: Hymn of Hate. *Datamation*, April 1 1971, p.31. Saved, for example, at OCC4CCC.HACKLIB and https://cahighways.org/wordpress/?tag=compusaur

[Wulf and Shaw 1973 Global] William Wulf and Mary Shaw. 1973. Global variable considered harmful. *SIGPLAN Not.* 8, 2 (February 1973), 28–34 [Shaw 1980 Abstr] Mary Shaw. The Impact of Abstraction Concerns on Modern Programming Languages. *Proc. IEEE special issue on Software Engineering*, 68, 9 (September 1980), pp.1119-1130 (invited).

[Shaw 1981 Good] Mary Shaw. When Is 'Good' Enough?: Evaluating and Selecting Software Metrics. In *Software Metrics: An Analysis and Evaluation*, (A. Perlis, F. Sayward, and M. Shaw, eds), MIT Press, 1981, pp.251-262.

[Shaw 1986 IO] Mary Shaw. An input-output model for interactive systems. *Proc. CHI '86: Conference on Human Factors in Computing Systems*, ACM SIGCHI, April 1986, pp.261–273.

[Shaw 1988 Abstr] Mary Shaw. Toward higher-level abstractions for software systems. *Proc. Tercer Simposio Internacional del Conocimiento y su Ingenieria*, October 1988 (printed by Rank Xerox) (invited), pp.55–61. Reprinted in *Data and Knowledge Engineering*, 5 (1990) pp.119–128. Revised and more accessible version, Larger-scale systems require higher-level abstractions, in *Proc. Fifth Int'l Workshop on Software Specification and Design*, Pittsburgh, May 1989, published as *Software Engineering Notes*, 14, 3, May 1989, pp.143–146.

[Shaw 1989 NextPL] Mary Shaw. Maybe your next programming language shouldn't be a programming language. *Scaling Up: A Research Agenda for Software Engineering*, National Academy Press, 1989, pp. 75-82.

[Lane 1990 Design Space] Thomas G. Lane. *User Interface Software Structures.* PhD thesis, Carnegie Mellon University, May 1990. Somewhat more accessible are two technical reports, *Studying Software Architecture Through Design Spaces and Rules*, CMU/SEI-90-TR-18 and CMU-CS-90-175, and *Design Space and Design Rules for User Interface Software Architecture*, CMU/SEI-90-TR-22 and CMU-CS-90-176.

[Shaw 1990 ArchLang] Mary Shaw. Elements of a design language for software architecture. Position paper for *IEEE Design Automation Workshop*, January 1990.

[Shaw 1991 Idiom] Mary Shaw. Heterogeneous design idioms for software architecture. *Proc. Sixth Int'l Workshop on Software Specification and Design*, IEEE, October 1991, pp.158–165.

[Bovik 1993 Perpetual] Harry Q. Bovik. Professional Student's Strategy for Perpetual Funding at CMU. *Carnegie Mellon University Computer Science Technical Report* CMU-CS-93-000, May 1993.

[Shaw 1993 Assy] Mary Shaw. Procedure calls are the assembly language of software interconnection: Connectors deserve first-class status. In D.A. Lamb (ed) *Studies of Software Design*, Proc of a 1993 Workshop, Lecture Notes in Computer Science No 1078, Springer Verlag 1996, pp.17–32.

[Shaw 1994 PLoP] Mary Shaw. Patterns for software architectures. First Annual Conference on the Pattern Languages of Programming, August 1994. In *Pattern Languages of Program Design*, vol 1, James Coplien and Douglas Schmidt (eds), Addison-Wesley 1995, pp. 453–462.

[Shaw 1995 Style] Mary Shaw, Comparing architectural design styles, *IEEE Software*, vol. 12, no. 6, pp. 27–41, Nov. 1995. doi: 10.1109/52.469758 (clearer version at http://www.cs.cmu.edu/~Compose/makingchoices.pdf)

[Shaw 1995 Control] Mary Shaw. Beyond objects: a software design paradigm based on process control. *ACM Software Engineering Notes*, 20, 1, Jan 1995. https://doi.org/10.1145/225907.225911

[Shaw et al 1995 ModelProb]. Mary Shaw, David Garlan, Robert Allen, Dan Klein, John Ockerbloom, Curtis Scott, and Marco Schumacher. *Candidate Model Problems in Software Architecture.* Working paper, version 1.3 January 1995. https://www.cs.cmu.edu/afs/cs/project/vit/ftp/pdf/ModPrb1-3.pdf Online version archived at https://web.archive.org/web/20071030070927/http://www.cs.cmu.edu:80/People/ModProb/ October 30 2007.

[Shaw 1995 PLoP] Mary Shaw. Some patterns for software architectures. Second Annual Conference on the Pattern Languages of Programming, 1995. In *Pattern Languages of Program Design, vol 2*, John M Vlissides, James O Coplien, and Norman L Kerth (eds), Addison Wesley 1995, pp. 255–269.

[Shaw 1996 SIS] Mary Shaw. Software architectures for shared information systems. In *Mind Matters: Contributions to Cognitive and Computer Science in Honor of Allen Newell.* Lawrence Erlbaum, 1996, pp.219-251.

[Shaw and Garlan 1996 SwArch] Mary Shaw and David Garlan. *Software architecture: Perspectives on an emerging discipline.* Prentice-Hall 1996.

[Shaw 1996 Truth] Mary Shaw, Truth vs. knowledge: the difference between what a component does and what we know it does, *Proceedings of the 8th International Workshop on Software Specification and Design*, Schloss Velen, Germany, 1996, pp. 181–185. Doi: 10.1109/IWSSD.1996.501165

[Shaw and Clements 1997 Boxology] Mary Shaw and Paul Clements. A field guide to boxology: Preliminary classification of architectural styles for software systems. *Proc. COMPSAC97, 21st Int'l Computer Software and Applications Conference*, August 1997, pp.6–13. doi: 10.1109/CMPSAC.1997.624691 .

[Butler et al 2000 GoodBad] Shawn Butler, Somesh Jha, and Mary Shaw. When good models meet bad data: Applying quantitative economic models to qualitative engineering judgments. Position paper, *Second Workshop on Economics-Driven Software Engineering Research (EDSER-2)*, May 2000. [nonarchival]

[Raz and Shaw 2000 SuffCor] Orna Raz and Mary Shaw. An approach to preserving sufficient correctness in open resource coalitions. *10th International Workshop on Software Specification and Design (IWSSD-10)*, Nov 2000, pp. 159–170.

[Shaw 2000 Ed] Mary Shaw. Software engineering education: A roadmap. *The Future of Software Engineering*, special volume of ICSE 2000, International Conference on Software Engineering, June 2000, pp. 373–380.

[Shaw 2000 Homeostasis] Mary Shaw. Sufficient correctness and homeostasis in open resource coalitions: How much can you trust your software system? *Proc 4th International Software Architecture Workshop (ISAW-4)*, May 2000.

[Shaw 2002 Everyday] Mary Shaw. Everyday dependability for everyday needs. *Supp Proc 13th International Symposium on Software Reliability Engineering*, pp.7–11, November 2002 (invited keynote).

[Shaw 2002 Heal] Mary Shaw. Self-healing: softening precision to avoid brittleness. In *Proceedings of the First Workshop on Self-healing Systems (WOSS '02)*. Association for Computing Machinery, 111–114.

[Poladian et al 2003 TimeMon] Vahe Poladian, Shawn A. Butler, Mary Shaw, and David Garlan. Time is not money: The case for multi-dimensional accounting in value-based software engineering. Position paper for *Fifth Workshop on Economics-Driven Software Research (EDSER-5)*, at the 25th Int'l Conf on Software Engineering, 2003.

[Shaw 2003 Writing] Mary Shaw, Writing good software engineering research papers*, Proc 25th International Conference on Software Engineering*, 2003. Portland, OR, USA, 2003, pp. 726-736.

[Scaffidi et al 2005 Number] Christopher Scaffidi, Mary Shaw and Brad Myers, Estimating the numbers of end users and end user programmers. *2005 IEEE Symposium on Visual Languages and Human-Centric Computing (VL/HCC'05),* Dallas, TX, USA, 2005, pp. 207-214, doi: 10.1109/VLHCC.2005.34.

[Shaw 2005 Spark] Mary Shaw, Sparking research ideas from the friction between doctrine and reality, *5th Working IEEE/IFIP Conference on Software Architecture (WICSA'05)*, Pittsburgh, PA, USA, 2005, pp. 11–16. doi: 10.1109/WICSA.2005.67

[Scaffidi et al 2006 Features] Christopher Scaffidi, Amy Ko, Brad Myers and Mary Shaw, Dimensions characterizing programming feature usage by information workers. *Visual Languages and Human-Centric Computing (VL/HCC'06),* 2006, pp. 59-64, doi: 10.1109/VLHCC.2006.21.

[Scaffidi and Shaw 2007 Confid] Christopher Scaffidi and Mary Shaw. Toward a calculus of confidence, 2007 *First International Workshop on the Economics of Software and Computation*, Minneapolis, MN, 2007, pp. 7–7.

[Scaffidi and Shaw 2007 Cred] Christopher Scaffidi and Mary Shaw. Developing confidence in software through credentials and low-ceremony evidence. *Int'l Workshop on Living with Uncertainties (IWLU'07),* co-located with the 23rd IEEE/ACM International Conference on Automated Software Engineering (ASE 2007), Atlanta, GA, November 2007.

[Shaw 2007 Depend] Mary Shaw. Strategies for achieving dependability in coalitions of systems. Position paper for *Dagstuhl Seminar 07031 on Software Dependability Engineering*, Schloss Dagstuhl, Germany, January 2007 [revision of 06ShRobust]

[Müller et al 2008 Control] Hausi Müller, Mauro Pezzè, and Mary Shaw. Visibility of control in adaptive systems. In *Proceedings of the 2nd International Workshop on Ultra-large-scale Software-intensive Systems (ULSSIS '08),* 2008. Association for Computing Machinery, New York, NY, USA, 23–26.

[Scaffidi et al 2008 Topes] Christopher Scaffidi, Brad Myers, and Mary Shaw. Topes: Reusable abstractions for validating data. *Proc International Conference on Software Engineering,,* 2008. 1–10.

[Bovik 2009 MIP] Harry Q. Bovik. Most Influential Paper from $2^{2^{2^{2^0}}}$ years ago. *Proc SIGBOVIK 2009*, April 5 2009, pp.1-9.. https://sigbovik.org/2009/proceedings.pdf

[Fidget and Nowhey 2011 Holiday] Tempus Fidget and Ukidding Nowhey. The Holiday Coverage Problem: the ultimate approach to deadline avoidance. *Proc SIGBOVIK 2011*, April 1 2011, pp.7-13, http://sigbovik.org/2011/proceedings.pdf

[Ko et al 2011 EUSE] Amy J. Ko, Robin Abraham, Laura Beckwith, Alan Blackwell, Margaret Burnett, Martin Erwig, Chris Scaffidi, Joseph Lawrance, Henry Lieberman, Brad Myers, Mary Beth Rosson, Gregg Rothermel, Mary Shaw, and Susan Wiedenbeck. 2011. The state of the art in end-user software engineering. *ACM Comput. Surv.* 43, 3, Article 21 (April 2011), 44 pages. https://doi.org/10.1145/1922649.1922658

[Shaw 2012 design] Mary Shaw. The role of design spaces. IEEE Software, vol. 29, no. 1, pp. 46-50, Jan.-Feb. 2012, doi: 10.1109/MS.2011.121.

[Shaw 2013 design] Mary Shaw. The role of design spaces in guiding a software design. In Software Designers in Action: a Human-Centric Look at Design Work. Marian Petre and André van der Hoek (ed). Chapman and Hall/CRC, 2013.

[Shaw 2016 Control] Mary Shaw. What can control theory teach us about designing cyber-physical systems? *IEEE World Forum on Internet of Things* Dec 2016. (video) https://ieeetv.ieee.org/mary-shaw-control-theory-and-designing-cyber-physical-systems-wf-iot-2016

[Litoiu et al 2017 CtlThy] Marin Litoiu, Mary Shaw, Gabriel Tamura, Norha M Villegas, Hausi A. Mueller, Holger Giese, Romain Rouvoy, and Eric Rutten. What can control theory teach us about assurances in self-adaptive software systems? In R. de Lemos et al (eds) *Software Engineering for Self-Adaptive Systems III, Assurances*. Lecture Notes in Computer Science vol 9640, Springer, 2017, pp 90–134. doi:10.1007/978-3-319-74183-3_4

[Le Goues 2018 Bridge] Claire Le Goues, Ciera Jaspan, Ipek Ozkaya, Mary Shaw, and Kathryn Stolee. Bridging the gap: From research to practical advice, in *IEEE Software*, vol. 35, no. 5, pp. 50–57, September/October 2018].

[Klawe et al 2020 EBS] Maria Klawe, Daniel V Klein, D K Fackler, Mary Shaw, Michael Ancas, Augie Fackler, Sarah R Allen, and Harry Q Bovik. Erdös-Bacon-Sabbath numbers—reductio ad absurdum. *Proc SIGBOVIK 2020*, April 1 2020, p.341. http://sigbovik.org/2020/proceedings.pdf

[Diogenes 2021 Rankings] Diogenes. Winning the rankings game: A new, wonderful, truly superior CS ranking. *Proc SIGBOVIK 2021*, April 1 2021, pp.158–162. http://sigbovik.org/2021/proceedings.pdf

[Shaw 2021 Myth] Mary Shaw. Myths and mythconceptions: what does it mean to be a programming language, anyhow? *Proc. ACM Program. Lang.* 4, HOPL, Article 234 (June 2020), 44 pages. doi: 10.1145/3480947

[Shaw and Petre 2024 Space] Mary Shaw and Marian Petre. Design spaces and how software designers use them: a sampler. Designing '24: 2024 International Workshop on Designing Software Proceedings. doi: 10.1145/3643660.3643941